\documentclass[12pt]{spieman}  
\usepackage{amsmath,amsfonts,amssymb}
\usepackage{graphicx}
\usepackage{setspace}
\usepackage{tocloft}
\usepackage{lineno}

\usepackage{xcolor}
\newcommand\todo[1]{{#1}}

\title{The Wide Area Linear Optical Polarimeter Control Software}

\author[a,b*]{John A. Kypriotakis}
\author[c]{Bhushan Joshi}
\author[a,b]{Dmitry Blinov}
\author[a,b]{Sebastian Kiehlmann}
\author[c,j]{Ramya M. Anche}
\author[a]{Ioannis Liodakis}
\author[a]{Myrto Falalaki}
\author[h]{Tuhin Ghosh}
\author[g]{Eirik Gjerløw}
\author[c,d]{Siddharth Maharana}
\author[a,b]{Nikolaos Mandarakas}
\author[f]{Georgia V. Panopoulou}
\author[a,b]{Katerina Papadaki}
\author[a,b]{Vasiliki Pavlidou}
\author[e]{Timothy J. Pearson}
\author[k]{Vincent Pelgrims}
\author[d,i]{Stephen B. Potter}
\author[c]{Chaitanya V. Rajarshi}
\author[c,a,e]{A. N. Ramaprakash}
\author[e,l]{Anthony C. S. Readhead}
\author[m]{Raphael Skalidis}
\author[a,b]{Konstantinos Tassis}
\affil[a]{Institute of Astrophysics, Foundation for Research and Technology-Hellas, Voutes, 70013 Heraklion, Greece}
\affil[b]{Department of Physics, University of Crete, Voutes, 70013 Heraklion, Greece}
\affil[c]{Inter-University Centre for Astronomy and Astrophysics, Post bag 4, Ganeshkhind, Pune, 411007, India}
\affil[d]{South African Astronomical Observatory, PO Box 9, Observatory, 7935, Cape Town, South Africa}
\affil[e]{Cahill Center for Astronomy and Astrophysics, California Institute of Technology, Pasadena, CA, 91125, USA}
\affil[f]{Department of Space, Earth and Environment, Chalmers University of Technology, 412 93, Gothenburg, Sweden}
\affil[g]{Institute of Theoretical Astrophysics, University of Oslo, P.O. Box 1029 Blindern, NO-0315 Oslo, Norway}
\affil[h]{School of Physical Sciences, National Institute of Science Education and Research, HBNI, Jatni 752050, Odisha, India}
\affil[i]{Department of Physics, University of Johannesburg, PO Box 524, Auckland Park 2006, South Africa}
\affil[j]{Steward Observatory, University of Arizona, Tucson, Arizona, 85721, USA}
\affil[k]{Universit\'e Libre de Bruxelles, Science Faculty CP230, B-1050 Brussels, Belgium}
\affil[l]{Owens Valley Radio Observatory, California Institute of Technology, MC 249-17, Pasadena, CA 91125, USA}
\affil[m]{Hubble Fellow, TAPIR, California Institute of Technology, MC 350-17, Pasadena, CA 91125, USA}

\cftpagenumbersoff{figure}
\cftpagenumbersoff{table} 
\begin{document} 
\maketitle

\begin{abstract}
The WALOPControl software is designed to facilitate comprehensive control and operation of the WALOP (Wide Area Linear Optical Polarimeter) polarimeters, ensuring safe and concurrent management of various instrument components and functionalities. This software encompasses several critical requirements, including control of the filter wheel, calibration half-wave plate, calibration polarizer, guider positioning, focusers, and 4 concurrent CCD cameras. It also manages the host telescope and dome operations while logging operational parameters, user commands, and environmental conditions for troubleshooting and stability. It provides a user-friendly graphical user interface, secure access control, a notification system for errors, and a modular configuration for troubleshooting are integral to the software’s architecture. It is accessible over the internet with the backend developed using NodeJS and ExpressJS, featuring a RESTful API that interacts with a MongoDB database, facilitating real-time status updates and data logging. The frontend utilizs the React.JS framework, with Redux for state management and Material UI for the graphical components. The system also allows for automatic observations based on user-defined schedules. A Continuous Integration and Continuous Deployment (CI/CD) pipeline ensures the software's reliability through automated testing and streamlined deployment. The WALOPControl software is a key component of the PASIPHAE (Polar-Areas Stellar Imaging in Polarimetry High Accuracy Experiment) project, which aims to study the dust and magnetic field of the Milky Way by observing the polarization of starlight.
\end{abstract}


{\noindent \footnotesize\textbf{*}J.A.K.,  \linkable{ikypriot@physics.uoc.gr} }

\begin{spacing}{2}   

\section{Introduction}\label{sect:intro}  
PASIPHAE (Polar-Areas Stellar Imaging in Polarimetry High Accuracy Experiment)\cite{pasiwhite} is a collaboration between the Institute of Astrophysics of the Foundation for Research and Technology-Hellas (IA-FORTH), the South African Astronomical Observatory (SAAO), the Inter-Univercity Centre for Astronomy and Astrophysics (IUCAA), the California Institute of Technology (Caltech) and the University of Oslo (UoO). The project aims to study the dust and magnetic field of the Milky Way by observing the polarization of starlight. The project will use 2 novel polarimeters, WALOPs (Wide Area Linear Optical Polarimeters)\cite{walopn, walops}, which will be mounted on the 1.3m telescope at the Skinakas Observatory in Crete, Greece and the Elizabeth 1m telescope of SAAO. The WALOPControl software is a web application that controls the operation of the WALOP polarimeters. The software is designed to be user-friendly and to provide a simple interface for the user to control the polarimeters and monitor their operation, while keeping extensive logging in its database and allowing for custom applications to run in conjunction to it. Furthermore, it is able to operate in a fully remote manner, with options to shut down in unexpected events such as bad weather.

The software is written in NodeJS\cite{ryan_dahl_nodejs_2009} and uses the React web framework\cite{react} to create a web application that can be accessed from any web browser. \todo{The backend of the software consists of about 20000 lines of code at the time of writting, while the frontend of about 45000 lines of code (including boilerplate code). A single developer/system administrator has been working on all aspects of the software for about 4 years, with the help of other developers for specific tasks (e.g. End-To-End testing - Section \ref{sect:test}).}

\todo{The completed software is set to be deployed in the Skinakas Observatory in Crete, Greece and the SAAO in South Africa, at the time of each instrument's commissioning. Nevertheless, as we will document below, extensive testing of its functionality has been carried out in the past years, using a testbed setup at the Inter-University Centre for Astronomy and Astrophysics in Pune, India. This testing helped us identify and fix several bugs, as well as to improve the overall functionality of the software.}

In this document we present the overall software ecosystem of the PASIPHAE project (Section \ref{sect:ecosystem}), the requirements of the WALOPControl software (Section \ref{sect:reqs}), and its architecture (Sections \ref{sect:back}, \ref{sect:front} and \ref{sect:db}).

\section{The PASIPHAE Software Ecosystem}\label{sect:ecosystem}

The PASIPHAE software ecosystem consists of several software components that are used to control the operation of the WALOP polarimeter, to analyze the obtained data and plan the observations carried out for the purposes of the PASIPHAE survey. The main components of the software ecosystem are:

\begin{itemize}
    \item The WALOPControl software, which is a full-stack web application that controls the operation of the WALOP polarimeter and the focus of this paper.
    \item The WALOPAnalysis software, which is a Python package that is used to analyze the data collected by the WALOP polarimeter and the focus of a later publication.
    \item The WALOPObservationPlanner software, which is a Python package that is used to plan the observations carried out by the WALOP polarimeters and the focus of a later publication.
\end{itemize}

The WALOPControl software is the main software component of the PASIPHAE project and is used to control the operation of the WALOP polarimeters. \todo{As inspiration for our software, we used the control software and analysis pipeline of RoboPol\cite{robopip}. The RoboPol pipeline is written in Python and provides a locally-installed GUI (Graphical User Interface) for the user to control the polarimeter (and its host telescope), monitor its operation and local weather, automatically observe lists of targets, support for interrupts to the nightly schedule, and analyze the data coming from the instrument. For the control system of WALOP, we shifted to a web-based approach (thus our change in technologies and the choice to write the software anew), but provide the same functionality, while also allowing for custom applications to run in conjunction to it and a more modular and extensible approach overall.}

The WALOPAnalysis software is a Python\cite{python} package that is used to analyze the data collected by the WALOP polarimeter. The software is designed to be able to carry out accurate photometry and polarimetry on the data, as well as implementing proper calibration\cite{calibration}. The software is written in Python, uses the Matplotlib\cite{matplotlib}, Astropy\cite{astropy} and Daophot\cite{daophot} libraries for data visualization and analysis. \todo{While the developing teams of the 2 software components are different, the WALOPControl software is designed to be able to provide WALOPAnalysis with appropriately formatted databases and data, in order to facilitate its smooth operation. This was made possible by the tight collaboration between the 2 teams, which allowed us to identify the requirements of the WALOPAnalysis software and implement them in the WALOPControl software.}

The WALOPObservationPlanner software is a Python package that is used to plan the observations carried out by the WALOP polarimeter, mainly for the purposes of the PASIPHAE project. The software is designed to be user-friendly in planning both long term and day-to-day observations, providing the guiding stars and optimizing the observing strategy for a selected patch of the sky. The software is written in Python, uses the Matplotlib, Astropy and Astroplan\cite{astroplan} libraries for observation planning and visualization. \todo{The nightly plan created by the WALOPObservationPlanner software is used by the WALOPControl software to carry out the observations robotically. The plan is generated in JSON (JavaScript Object Notation)\cite{smith_introducing_2015} format, while the WALOPControl software is designed to be able to read the plan and carry out the observations accordingly with only supervision from a human observer.}

\begin{figure}[!ht]
\begin{center}
\includegraphics[width=0.8\textwidth]{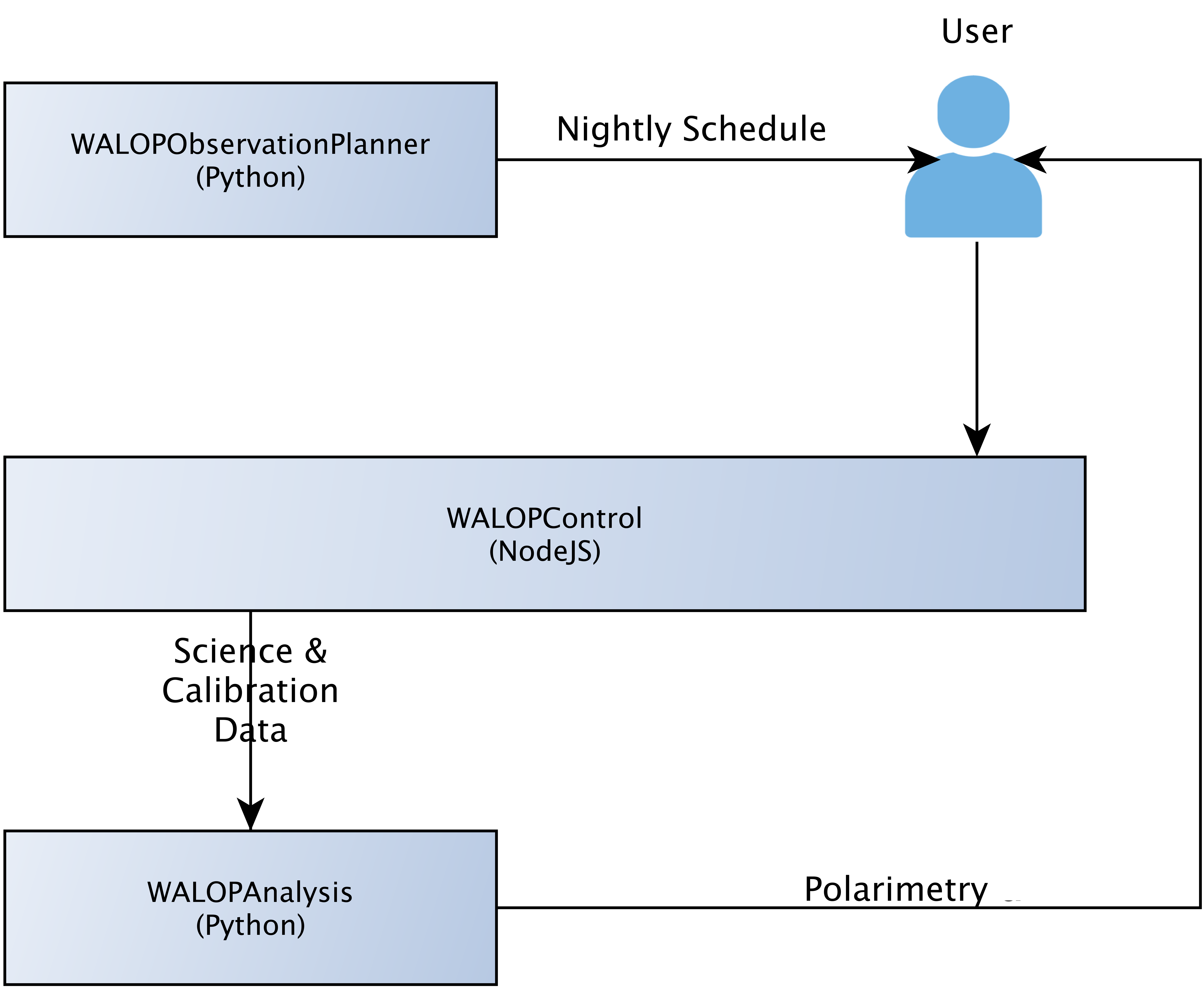}
\end{center}
\caption 
{ \label{fig:ecosystem}
The PASIPHAE software ecosystem. The WALOPControl software controls the operation of the WALOP polarimeter. The WALOPAnalysis software analyzes the data collected by the WALOP polarimeter. The WALOPObservationPlanner software plans the observations to be carried out by the WALOP polarimeter.} 
\end{figure}

Figure \ref{fig:ecosystem} shows the overall software ecosystem of the PASIPHAE project.

\section{The WALOP Polarimeter}\label{sect:walop}

The WALOP polarimeters\cite{walopn, walops} are novel polarimeters designed to study the dust and magnetic field of the Milky Way by observing the polarization of starlight. Each polarimeter consists of a collimator, a polarization optical array and 4 camera optical arrays. The collimator includes a filter wheel, a calibration half-wave plate (HWP), and a calibration polarizer. Each camera array terminates at a CCD (charge-coupled device) which moves perpendicularly to its optical axis for focusing. Each instrument includes a guider which can pivot around the science field of view (FoV) in a circular (in the case of WALOP-North) or rectilinear (in the case of WALOP-South) manner.

The filter wheel is used to select the wavelength of the light that is observed by the polarimeter. It can contain up to 4 filters and uses a stepper motor coupled with a platform on a lead screw (converting the stepper's motion from rotary to linear) to shift the appropriate filter into position.

The calibration HWP and polarizer are used to calibrate the polarimeter. They each use 2 stepper motors for their operation. One of them is coupled with a platform on a lead screw to shift the element in or out of the optical beam, while the other is coupled to a gear used to rotate the element around its optical axis. The rotation can be to a fixed angle or continuous.

The CCD cameras are used to collect the data that is observed by the polarimeter. They are the e2v model 231-84\cite{teledyne} controlled through an IUCAA Digital Sampler Array Controller (IDSAC)\cite{idsac} each. They are cooled to a temperature of -100$^{\circ}$C in order to reduce the dark current in the data.

The focusers are used to move the CCD cameras perpendicularly to their optical axis in order to focus the light that is observed by the polarimeter. They use a stepper motor coupled with a platform on a lead screw to move the CCD camera linearly.

\begin{figure}[!ht]
\begin{center}
\includegraphics[width=0.8\textwidth]{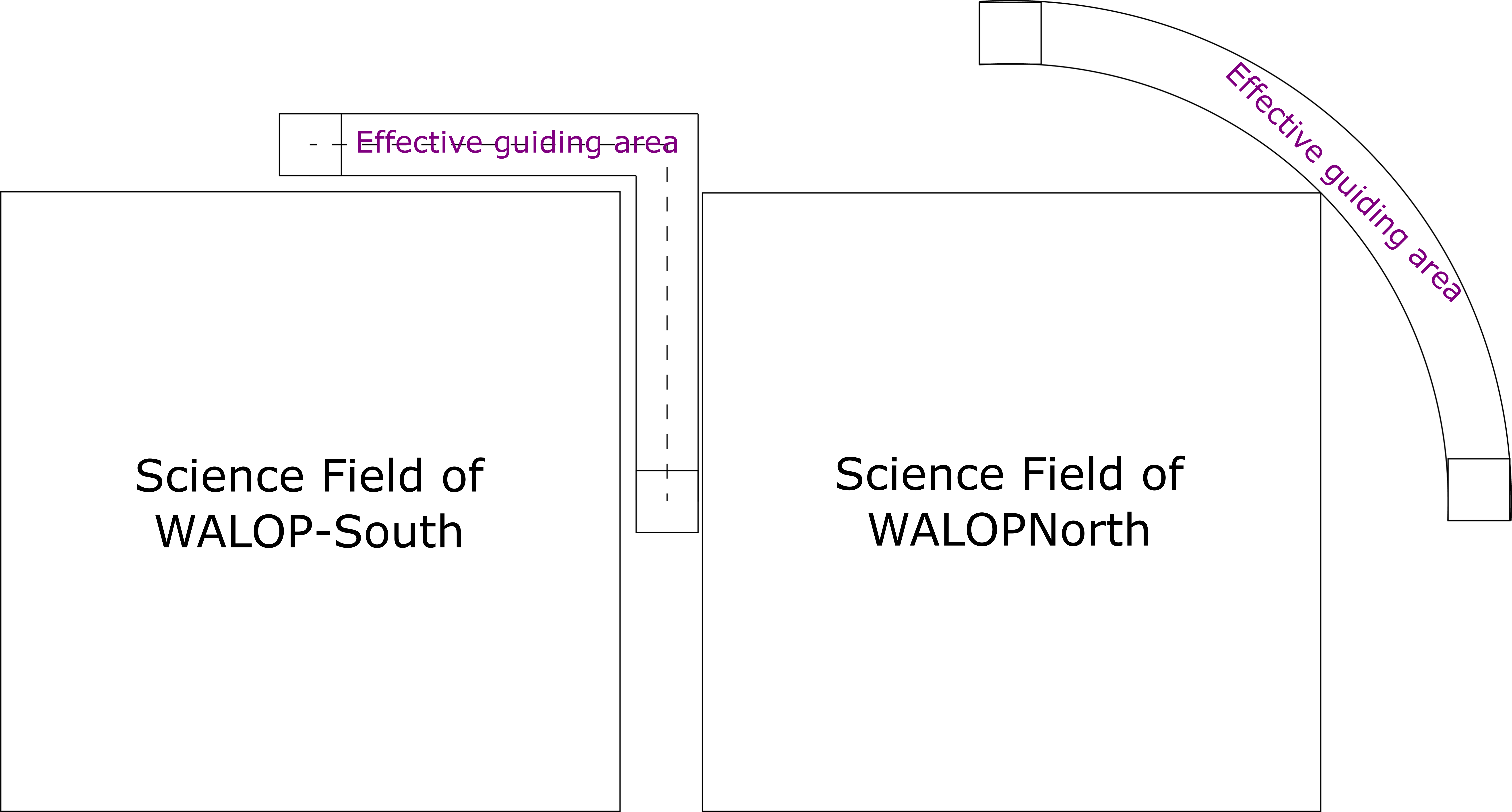}
\end{center}
\caption 
{ \label{fig:guider}
Comparison of a rectilinearly pivoting guider concept, used in WALOP-South (left) and a rotating guider concept, used in WALOP-North (right). The small squares in each case are the guider fields, and the big ones the science fields in each case (the rotating design's path has been truncated to $90^{\circ{}}$ for illustration purposes).} 
\end{figure}

The guider is for guiding the host telescope, reducing errors of pointing. In the case of WALOP-North it is a rotating guider, whereby a stepper motor gear-coupled to a platform rotates the entire guider array (pickup mirror, focusing optics and guider camera) around the collimator. In the case of WALOP-South it is a rectilinearly pivoting guider, using 2 stepper motors, each coupled to a lead screw (in perpendicular directions) and platform, to move the guider array in a rectilinear manner around colimator in an "L" shape. The movement of the guider's FoV around the science FoV is illustrated in Figure \ref{fig:guider}.

The control of all stepper motors is through a Motion Control Card (MCC - 1 for every 4 motors). All motion control card respond to commands over TCP/IP, as described in Appendix \ref{sect:ap_mcc}. All CCDs are controlled and read through an IDSAC, which responds to commands over USB 3.0\cite{usb_30_promoter_group_usb_2008}. Both the Skinakas 1.3m telescope and the Elizabeth 1m telescope are controlled through their respective control systems, which respond to commands over TCP/IP (Transmission Control Protocol/Internet Protocol). The weather station of Skinakas responds to requests conforming to the ASCOM Alpaca protocol\cite{ascom} over REST API\cite{rest} (Representational State Transfer Application Programming Interface). The weather station of SAAO responds to requests over non-Alpaca REST API. 

The controllables and telematics \todo{(in the context of this paper we will define telematics as information that is not necessarily controllable - e.g. the weather)} of the WALOP-North instrument are presented in Appendix \ref{sect:ap_north},  and the controllables and telematics of the WALOP-South instrument are presented in Appendix \ref{sect:ap_south}.

\section{WALOPControl Requirements}\label{sect:reqs}

The WALOPControl software has several requirements that need to be met in order to ensure the proper operation of the WALOP polarimeter. The main requirements of the software are:

\begin{itemize}
    \item Control the operation of the WALOP polarimeter, including the control of the filter wheel, the rotation of the calibration half-wave plate (HWP), the rotation of the calibration polarizer, the control of the guider positioning, the control of the instruments focusers, and the control of the CCD cameras. All operations should be done in a safe and concurrent manner.
    \item \todo{The control of all instrument motors should be accurate to the nearest $\mu{}m$ for all motors, even after repeat movements. This is to ensure that all components of the instrument are properly aligned and that the instrument can be used for accurate observations. Testing should be done to ensure that the accuracy of the movements is within the required limits.}
    \item \todo{Although time-domain astronomy is note expected to be carried out by the WALOP polarimeters, the software should be able to time the CCD to greatest extent. An arbitrary limit we set is for the CCDs to be timed within an order-of-magnitude of $10\mu{}s$, with the exposures being timed to the nearest $1\mu{}s$.}
    \item Control the host telescope (in the case of WALOPNorth) and the dome of the telescope, in order to ensure the proper and efficient operation of the instrument. \todo{This includes interfacing with the existing telescope control system (TCS) of both telescopes and their integrated safety systems.}
    \item Log the operational parameters of the instrument, as well as the user commands for troubleshooting, maintenance, and stability purposes. It should also be able to log the weather conditions \todo{(by interfacing with both observatories' weather system)} and the status of the host telescope and dome.
    \item Provide a simple interface for the user to control the operation of the instrument and monitor its operation.
    \item Run custom applications in conjunction to it, in order to provide customizable functionality to the user.
    \item Provide a secure login system that allows for different levels of access to the software.
    \item Provide a notification system that notifies the user and admins of any errors or warnings that occur during the operation of the WALOP polarimeter.
    \item Provide a modular configuration system that allows the admins to troubleshoot and customize the instrument control.
    \item Provide a data visualization system that allows the user to visualize the data collected by the WALOP polarimeter.
    \item Conduct automatic observations, based on a user-defined nightly schedule.
    \item \todo{Be integrable to the infrastructure existing or procurable at the observatories(Skinakas Observatory in Crete, Greece and SAAO in South Africa).}
    \item \todo{Even though the programming language and libraries used in the software are expected to be supported for at least 10 years (necessary for the PASIPHAE project's goals), the software should be designed to be easily maintainable and extensible, in order to ensure that it can be updated and maintained during its lifetime. The languages and technology used is not mandated by the PASIPHAE project, but rather chosen by the developers, in order to ensure that the software is fast, efficient and easy to use.}
\end{itemize}

\section{WALOPControl Backend}\label{sect:back}
The backend of the WALOPControl software is responsible for controlling the operation of the WALOP polarimeter, logging the operational parameters of the instrument, and providing a secure login system for the user. The backend is written in NodeJS and uses the Express web framework\cite{tj_holowaychuk_expressjs_2010} to create a RESTful API that can be accessed from the frontend. The choice of NodeJS and Express was made to ensure that the backend is fast and efficient, as well as to provide a simple and easy-to-use, frontend-agnostic API.

\begin{figure}[!ht]
\begin{center}
\includegraphics[width=0.9\textwidth]{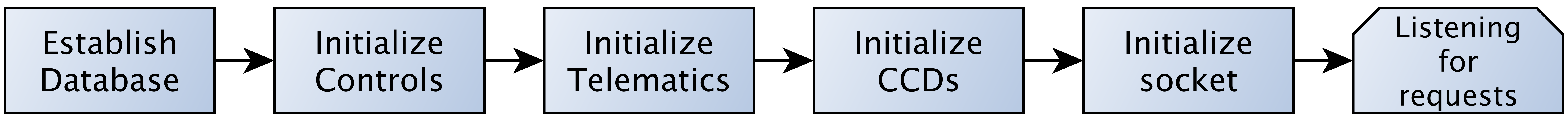}
\end{center}
\caption 
{ \label{fig:backend}
The lifecycle of the WALOPControl backend.} 
\end{figure}

The lifecycle of the backend is presented in Figure \ref{fig:backend}. The backend starts by establishing a proper connection to the database. WALOPControl will then initialize all instrument controls and telematics, will open a WebSocket\cite{websocket} and finally start the server and listen for incoming requests from the frontend (or other software). When a request is received, the backend processes it and sends a response back to the requestor.

\begin{figure}[!ht]
\begin{center}
\includegraphics[width=0.9\textwidth]{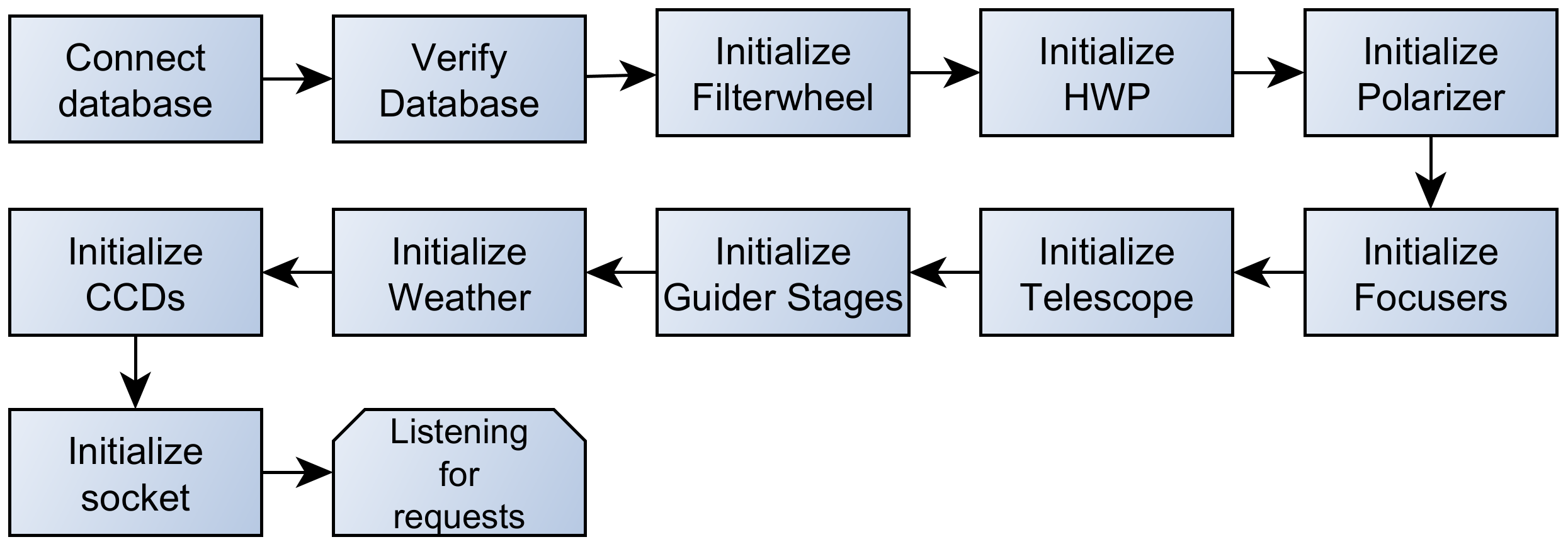}
\end{center}
\caption 
{ \label{fig:backend_init}
The initialization of the WALOPControl-backend, extending Figure \ref{fig:backend}.} 
\end{figure}

The series of initialization steps that the backend follows are shown in Figure \ref{fig:backend_init}. After the database is connected to and its integrity is verified, the controls (Filterwheel, calibration HWP, calibration Polarizer, Focusers, Guider Stages) are initialized, then the telematics (TCS, and Weather Station) are initialized. Finally, the CCD connection is initialized, a WebSocket is created and the backend starts the request handling process.

\todo{Since the startup procedure occurs only once per reboot of the software, which is not a frequent event, the initialization process is not time-critical. The initialization process is designed to be robust and to handle errors gracefully, ensuring that the backend can recover from any errors that occur during the initialization process. The above mentioned steps are therefore completed sequentially to allow for the engineer who is responsible for the instrument to monitor the initialization process and to intervene if necessary.}

\subsection{Establishing Database}\label{sect:db_init}
The first step in the initialization process is to establish a proper connection to the database. The database \todo{(MongoDB\cite{mongo} - detailed in Section \ref{sect:db})} is first connected to, using the mongoose\cite{mongoose} library. It is then verified that the following criteria are met by the database:
\begin{enumerate}
    \item Is accessible (read permission).
    \item Has not been altered since the previous WALOPControl-backend shutdown. An exception is made if the database was altered with justification in the appropriate database collection (Section \ref{sect:db}).
    \item Includes at least 1 user.
    \item Includes at least 1 admin.
    \item Can be written to (write permission).
\end{enumerate}

\subsection{Initialization of Instrument Controls}\label{sect:inst_init}
The next step is to initialize all instrument controls and telematics. This is done by sending a series of commands to the MCCs, in order to ensure that all instruments are in a safe and stable state. The procedure differs between the initialization of linear and rotary motors. The initialization of linear motors is done by homing the motor to a specified end-switch position and then moving to a valid position in the linear path, while the initialization of rotary motors is done by only homing the motor to a specified end-switch position. These procedures are repeated for all motors in the instrument, depending on their coupling (linear or rotary).

\subsubsection{Initialization of Linear Motors}\label{sect:lin_init}
\begin{figure}[!ht]
\begin{center}
\includegraphics[width=0.9\textwidth]{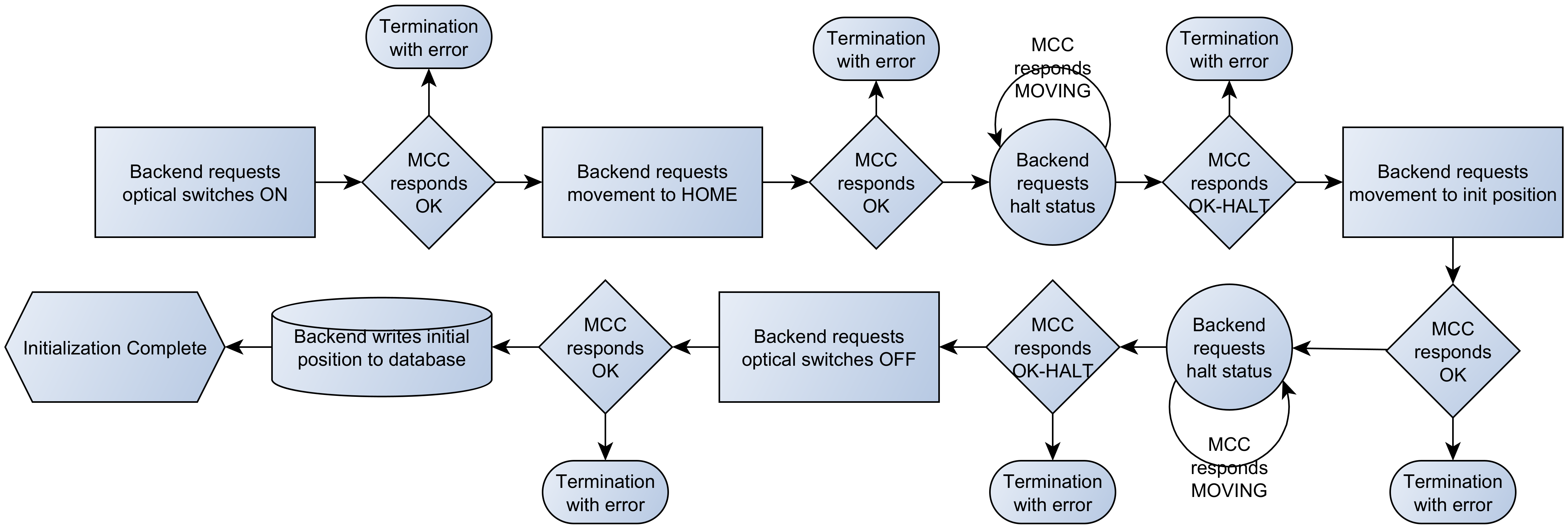}
\end{center}
\caption 
{ \label{fig:lin_init}
The initialization protocol for the linear motors of the WALOP instrument. It is carried out every time the software (re)starts for every linear motor of the instrument.} 
\end{figure}

The initialization of any linear motor is done by sending the SNON command to the MCC (for the motor in question) to turn on the optical switches, followed by the HOME command to home the motor, the SFIN command to move the motor to a valid position in the linear path and the SNOF command to turn the optical switches off. Finally, the position of the motor is logged in the database as the current position of the control. After every command, the response of the MCC is checked for errors and during every movement, the status of the motor is checked using the GMST command recursively until the motor halts. Upon any error, the initialization process is halted. Figure \ref{fig:lin_init} shows the initialization process of a linear motor. The usage of the SNON command is to ensure that the motor does not run away during the movement (checked by the limit switches) and that homing is successful (executed by the home switch). The SNOF command is used to ensure that the optical switches do not interfere with the observations (the light switches are off during observations).

\subsubsection{Initialization of Rotary Motors}\label{sect:rot_init}
\begin{figure}[!ht]
\begin{center}
\includegraphics[width=0.9\textwidth]{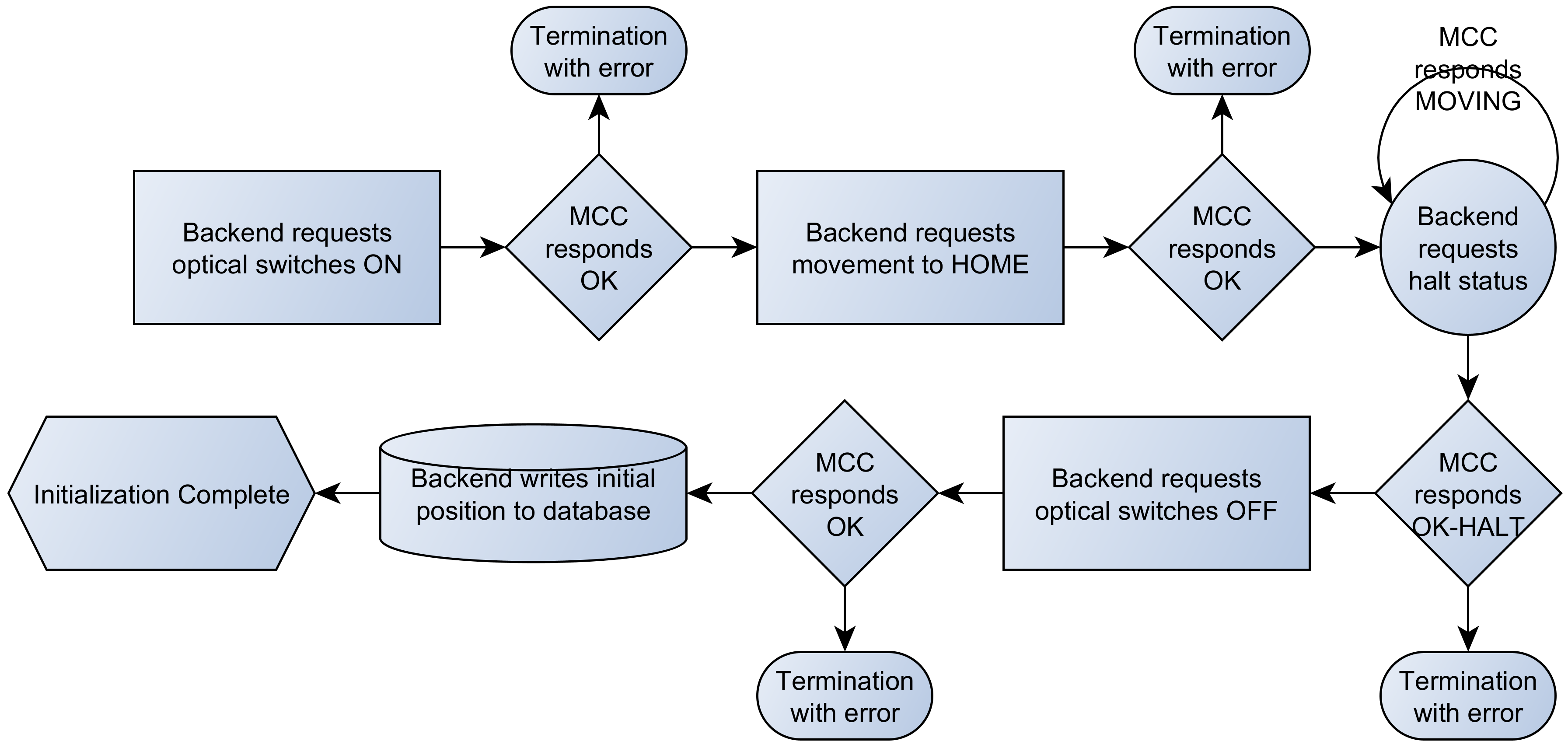}
\end{center}
\caption 
{ \label{fig:rot_init}
The initialization protocol for the rotary motors of the WALOP instrument. It is carried out every time the software (re)starts for every rotary motor of the instrument.} 
\end{figure}

The initialization of any rotary motor is done by sending the SNON command to the MCC (for the motor in question) to turn on the optical switches, followed by the HOME command to home the motor and the SNOF command to turn the optical switches off. Finally, the position of the motor is logged in the database as the current position of the control. After every command, the response of the MCC is checked for errors and during every movement, the status of the motor is checked using the GMST command recursively until the motor halts. Upon any error, the initialization process is halted. Figure \ref{fig:rot_init} shows the initialization process of a rotary motor.

\subsection{Initialization of Instrument Telematics}\label{sect:tel_init}
The next step is to initialize all instrument telematics. This is done by sending a series of commands to the respective interfaces, in order to ensure that all telematics are in a safe and stable state.

\subsubsection{Initialization of the Telescope Connection}\label{sect:tcs_init}
The initialization of the telescope connection is done by creating a TCP/IP socket object, binding it to the TCS IP and attaching it to the running server object for future use. The connected TCS is checked for proper time and coordinate refence settings. After that, a ``get status" command is issued to the TCS, following the communication protocol of the respective telescope control system. The response is checked for errors and the status of the telescope is logged in the database.

Having established the connection to the TCS, a CRON job is set up and attached to the server object, issuing a ``get status" command to the TCS and logging the status of the telescope in the database. The job will be ran every 1, 60, or 300 seconds, depending on whether the instrument is in slewing, observing or idle mode respectively (and the configuration of the software). The job is stopped when the software is shut down.

\subsubsection{Initialization of the Weather Station Connection}\label{sect:wea_init}
The initialization of the weather station connection is also done by creating a connection object (depending on the communication protocol of the respective weather station) and attaching it to the running server object for future use. The connected weather station is checked for proper time settings. After that, a ``get weather" command is issued, following the communication protocol of the respective weather station. The response is checked for errors and the weather is logged in the database.

Having established the connection to the weather station, a CRON job is set up and attached to the server object, issuing a ``get weather" command to the station and logging the weather in the database. The job will be ran every 1-5 minutes, depending on the configuration of the software. The job is stopped when the software is shut down.

\subsection{Initialization of the CCDs}\label{sect:ccd_init}
The WALOP polarimeter uses 4 CCD cameras to collect the data that is observed by the instrument. The CCDs are controlled via an IDSAC\cite{idsac} card (each). In order to communicate with the IDSAC cards, we need to utilize a USB 3.0 connection. Our software was designed to be accessible over the web and housed in an arbitrary server, which may not have direct access to the IDSAC cards. To overcome this limitation, we developed a separate software (called WALOPCamera), which is installed on a small-form-factor PC (SFFPC) within the WALOP instruments.

\subsubsection{The WALOPCamera Software}\label{sect:walopcam}
This PC is connected to the 4 onboard IDSAC cards. The WALOPCamera software is responsible for receiving commands (related to the CCD operations) from WALOPControl over TCP/IP and controlling the CCDs (according to those commands) via the IDSAC cards. The WALOPControl software is responsible for sending the commands to the WALOPCamera software and receiving the data from the CCDs. The WALOPCamera software is written in C and uses the IDSAC API library\cite{idsacapi} to communicate with the IDSAC cards. The choice of programming language was made to ensure the fastest possible communication between the IDSAC cards and the WALOPCamera software, therefore reducing the exposure delay between the 4 CCDs. The commands available to the WALOPCamera software user are listed in Appendix \ref{sect:ap_walopcamera}.

\subsubsection{CCD Initialization}\label{sect:init_ccd}
The initialization of the WALOPCamera connection is done by creating a TCP/IP socket object, binding it to the SFFPC's IP and attaching it to the running server object for future use. The connected WALOPCamera software is issued a SETT command, setting the time of the SFFPC. The GAIN, FREQ and TEMP commands are issued in accordance with the WALOPControl configuration for the default CCD settings. The SUBF command is issued to set the subframe of the CCDs to the whole CCD. The BINN command is issued to set the binning of the CCDs in accordance with the WALOPControl configuration. The HEAD command is issued recursively setting all headers to default (exposure-independent) values.

Having set up the CCDs, the EXPO command is used to set the exposure time to 0 (bias observations). The MULT command is used to set exposure block to 10 exposures. The STAT command is used recursively to check that the CCDs have cooled to the aforeset temperature. The STRT command is issued to start the acquisition of the 10 bias frames. WALOPCamera takes care of exposing the CCDs and saving the bias frames as FITS files\cite{fits}. The STAT command is used recursively to check that the exposures have concluded. When so, the bias frames are recorded in the database as observed and the CCDs are ready for operation.

\subsection{Final Steps}\label{sect:final}
After the initialization of all instrument controls and telematics, the backend opens a WebSocket (that is used for state updates on the client side) and starts the server, listening for incoming requests from the frontend (or other piece of software that may want to use the backend services).

\subsection{Authentication}\label{sect:auth}
The backend provides a secure login system that allows for different levels of access to the software. The login process is done by checking the provided username and password against the database. If the credentials are correct, a JWT (JSON Web Token)\cite{jwt} is generated and sent back to the user. This is used for ``Bearer Authentication", whereby the JWT is then used to authenticate the user for future requests. The JWT is valid for a certain amount of time (configurable by the admin) and is stored in the user's local storage. The JWT is then sent back to the backend with every request in an ``Authentication" header entry, in order to authenticate the user.

\subsection{Request Handling}\label{sect:req_handling}
When a request is received by the backend, it is processed and a response is sent back to the requestor. The request is first checked for errors (using express.js compatible middleware - Section \ref{sect:middleware}) and then processed according to the request type (by the request controller - Section \ref{sect:controllers}). The response is then sent back to the requestor, according to the request type. It is important to note that all requests are handled concurrently and independently of each other, in order to ensure that the backend is fast, concurrent and efficient. Only concurrent requests to the same controllable are ignored (e.g. sending multiple ``Change Filter" commands at the same time). This feature is among the ones tested in in the End-To-End Testing (Section \ref{sect:test}).

\subsubsection{Middleware}\label{sect:middleware}
Middleware are functions used to check for errors in requests, and enrich the request objects by adding or pre-processing the information included in the original request. Each middleware is usable by multiple different controllers and has the power to either end the request processing procedure by returning an error response to the requestor, or pass the request forward to the next middleware or  the request controller (if no errors were detected). WALOPControl-backend offers the following middleware:
\begin{enumerate}
    \item\label{middle:loggedin} ``User Logged In": Checks if a user is logged in, via the provided Bearer's JWT. Adds the user's details from the database to the request (censoring the password and other confidential entries). Responds \verb|401 - Unauthorized| if JWT is missing or invalid/expired. Used in all routes except signup.
    \item\label{middle:approved} ``User Approved": Checks if user (provided by Middleware \ref{middle:loggedin}) is approved to observe. Responds \verb|401 - Unauthorized| if not. Used in all routes, except signup.
    \item\label{middle:admin} ``Is Admin": Checks if user (provided by Middleware \ref{middle:loggedin}) is an administrator. Responds \verb|401 - Unauthorized| if not. Used in administrator-only routes.
    \item\label{middle:nosched} ``No Schedule Running": Checks if no schedule is running in automatic observation mode. Responds with \verb|403 - Forbidden| if it is. Used in endpoints that would intervene with automatic observations.
    \item\label{middle:filternotmoving} ``Filter-wheel Not Moving": Checks if the filter-wheel is not moving. Responds with \verb|403 - Forbidden| if it is. Used in endpoints that will change the filter.
    \item\label{middle:filterok} ``Filter OK": Checks if the filter provided in the request is accessible. Responds with \verb|404 - Not Found| if not. Used in endpoints that will change the filter.
    \item\label{middle:hwpnotmoving} ``HWP Not Moving": Checks if the calibration HWP is not moving. Responds with \verb|403 - Forbidden| if it is. Used in endpoints that will change the status of the calibrator HWP.
    \item\label{middle:hwpok} ``HWP OK": Checks if the HWP configuration (in/out of beam, angle, continuous motion parameters) provided in the request is accessible. Responds with \verb|404 - Not Found| if not. Used in endpoints that will change the status of the HWP.
    \item\label{middle:polnotmoving} ``Polarizer Not Moving": Checks if the calibration polarizer is not moving. Responds with \verb|403 - Forbidden| if it is. Used in endpoints that will change the status of the calibrator polarizer.
    \item\label{middle:polok} ``Polarizer O"K": Checks if the polarizer configuration (in/out of beam, angle, continuous motion parameters) provided in the request is accessible. Responds with \verb|404 - Not Found| if not. Used in endpoints that will change the status of the polarizer.
    \item\label{middle:focnotmoving} ``Focuser Not Moving": Checks if the provided focuser is not moving. Responds with \verb|403 - Forbidden| if it is. Used in endpoints that will change the status of the focuser.
    \item\label{middle:focok} ``Focuser OK": Checks if the focuser position provided in the request is accessible. Responds with \verb|404 - Not Found| if not. Used in endpoints that will change the status of the focuser.
    \item\label{middle:guidernotmoving} ``Guider Not Moving": Checks if the guider's positioning is not moving. Responds with \verb|403 - Forbidden| if it is. Used in endpoints that will change the positioning of the guider.
    \item\label{middle:guiderok} ``Guider OK": Checks if the guider positioning provided in the request is accessible. Responds with \verb|404 - Not Found| if not. Used in endpoints that will change the status of the guider positioning.
    \item\label{middle:telescopenotslew} ``Telescope Not Slewing": Checks if the telescope is not slewing. Responds with \verb|403 - Forbidden| if it is. Prevents multiple overlapping slew commands. Used in endpoints that will request a telescope slew.
    \item\label{middle:telescopeslewstop} ``Telescope Slewing Stoppable": Checks if the telescope is slewing with a stoppable slew. Responds with \verb|403 - Forbidden| if it is not. Prevents stop slew commands on a halted telescope or during critical pointing. Used in endpoints that will request a telescope slew stop.
    \item\label{middle:targetavail} ``Telescope Target Available": Checks if the target (either by name, equatorial, or horizontal coordinates) provided in the request is accessible. Responds with \verb|403 - Forbidden| if not. Responds with \verb|404 - Not Found| if the target does not exist. Used in endpoints that request telescope slewing.
    \item\label{middle:notobserving} ``CCD Not Observing": Checks if the CCD is not observing. Responds with \verb|403 - Forbidden| if it is. Prevents multiple overlapping observe commands. Used in endpoints that will request CCD observations.
    \item\label{middle:observingstop} ``CCD Observation Stoppable": Checks if the CCD is observing with a stoppable observation. Responds with \verb|403 - Forbidden| if it is not. Prevents stop observation commands on dormant CCD or during critical observations. Used in endpoints that will request a CCD observation stop.
    \item\label{middle:observationpossible} ``Observation Possible": Checks if the observation parameters (number of exposures, exposure time) provided in the request are possible. Responds with \verb|403 - Forbidden| if not. Used in endpoints that request CCD observations.
\end{enumerate}

\subsubsection{Controllers}\label{sect:controllers}

Controllers are functions used to process the requests that are received by the backend. Each controller is responsible for a specific type of request and is used to process the request and send a response back to the requestor. The controllers available in WALOPControl-backend are presented in Table \ref{tab:controllers}.

\begin{table}[ht]
\caption{Controllers used in the WALOPControl-backend.} 
\label{tab:controllers}
\begin{center}       
\begin{tabular}{||||||c|c|c|c||||||} 
\hline\hline\hline
\hline\hline\hline
\rule[-1ex]{0pt}{3.5ex} Service & Control & Middlewares & Handling \\
\hline\hline\hline
\rule[-1ex]{0pt}{3.5ex}  Filter-wheel & Check Status & \ref{middle:loggedin}, \ref{middle:approved} & \ref{subsubsec:status} \\
\hline
\rule[-1ex]{0pt}{3.5ex}  Filter-wheel & Set Filter &  \ref{middle:loggedin}, \ref{middle:approved}, \ref{middle:nosched}, \ref{middle:filternotmoving}, \ref{middle:filterok} & \ref{subsubsec:linear_motor} \\
\hline
\rule[-1ex]{0pt}{3.5ex}  Calibration HWP & Check Status & \ref{middle:loggedin}, \ref{middle:approved} & \ref{subsubsec:status} \\
\hline
\rule[-1ex]{0pt}{3.5ex}  Calibration HWP & Set Position &  \ref{middle:loggedin}, \ref{middle:approved}, \ref{middle:nosched}, \ref{middle:hwpnotmoving}, \ref{middle:hwpok} & In/Out movement: \ref{subsubsec:linear_motor}, Rotation: \ref{subsubsec:rotary_motor} \\
\hline
\rule[-1ex]{0pt}{3.5ex}  Calibration polarizer & Check Status & \ref{middle:loggedin}, \ref{middle:approved} & \ref{subsubsec:status} \\
\hline
\rule[-1ex]{0pt}{3.5ex}  Calibration polarizer & Set Position &  \ref{middle:loggedin}, \ref{middle:approved}, \ref{middle:nosched}, \ref{middle:polnotmoving}, \ref{middle:polok} & In/Out movement: \ref{subsubsec:linear_motor}, Rotation: \ref{subsubsec:rotary_motor} \\
\hline
\rule[-1ex]{0pt}{3.5ex}  Each Focuser & Check Status & \ref{middle:loggedin}, \ref{middle:approved} & \ref{subsubsec:status} \\
\hline
\rule[-1ex]{0pt}{3.5ex}  Each Focuser & Set Focus Position &  \ref{middle:loggedin}, \ref{middle:approved}, \ref{middle:nosched}, \ref{middle:focnotmoving}, \ref{middle:focok} & \ref{subsubsec:linear_motor} \\
\hline
\rule[-1ex]{0pt}{3.5ex}  Guider & Check Status & \ref{middle:loggedin}, \ref{middle:approved} & \ref{subsubsec:status} \\
\hline
\rule[-1ex]{0pt}{3.5ex}  Guider & Change Positioning &  \ref{middle:loggedin}, \ref{middle:approved}, \ref{middle:nosched}, \ref{middle:guidernotmoving}, \ref{middle:guiderok} & \ref{subsubsec:linear_motor} \\
\hline
\rule[-1ex]{0pt}{3.5ex}  Telescope & Check Status & \ref{middle:loggedin}, \ref{middle:approved} & \ref{subsubsec:telescope} \\
\hline
\rule[-1ex]{0pt}{3.5ex}  Telescope & Set Next Slew Target &  \ref{middle:loggedin}, \ref{middle:approved}, \ref{middle:nosched}, \ref{middle:telescopenotslew}, \ref{middle:targetavail} & \ref{subsubsec:telescope} \\
\hline
\rule[-1ex]{0pt}{3.5ex}  Telescope & Slew to Next Target &  \ref{middle:loggedin}, \ref{middle:approved}, \ref{middle:nosched}, \ref{middle:telescopenotslew}, \ref{middle:targetavail} & \ref{subsubsec:telescope} \\
\hline
\rule[-1ex]{0pt}{3.5ex}  Telescope & Stop Slew &  \ref{middle:loggedin}, \ref{middle:approved}, \ref{middle:nosched}, \ref{middle:telescopeslewstop} & \ref{subsubsec:telescope} \\
\hline
\rule[-1ex]{0pt}{3.5ex}  Weather & Get Weather & \ref{middle:loggedin}, \ref{middle:approved} & \ref{subsubsec:weather} \\
\hline
\rule[-1ex]{0pt}{3.5ex}  CCDs & Get Status & \ref{middle:loggedin}, \ref{middle:approved} & \ref{subsubsec:camera} \\
\hline
\rule[-1ex]{0pt}{3.5ex}  CCDs & Expose & \ref{middle:loggedin}, \ref{middle:approved}, \ref{middle:nosched}, \ref{middle:notobserving}, \ref{middle:observationpossible} & \ref{subsubsec:camera} \\
\hline
\rule[-1ex]{0pt}{3.5ex}  CCDs & Stop Exposure & \ref{middle:loggedin}, \ref{middle:approved}, \ref{middle:nosched}, \ref{middle:observingstop} & \ref{subsubsec:camera} \\
\hline\hline\hline
\hline\hline\hline
\end{tabular}
\end{center}
\end{table}

\subsubsection{Status check for components controlled via MCC}\label{subsubsec:status}

When the API user requests the status of the filter-wheel, calibrator HWP, calibration polarizer, focusers, or guider positioning, WALOPControl-backend polls the database for the latest document in the respective collection (Section \ref{sect:db}), and responds to the user with \verb|200 - OK| and the contents of said document.

\subsubsection{Control of linear motors}\label{subsubsec:linear_motor}

All linear motors have a common control scheme. That means that the API endpoint is different for setting the filter-wheel position, the calibration HWP position in/out of the beam, the calibration polarizer position in/out of the beam, each focuser's position, and each guider motor position, but the logic behind it is the same. We will give below an example of a request to change the filter position.

\begin{figure}[!ht]
\begin{center}
\begin{tabular}{c}
\includegraphics[width=0.70\textwidth]{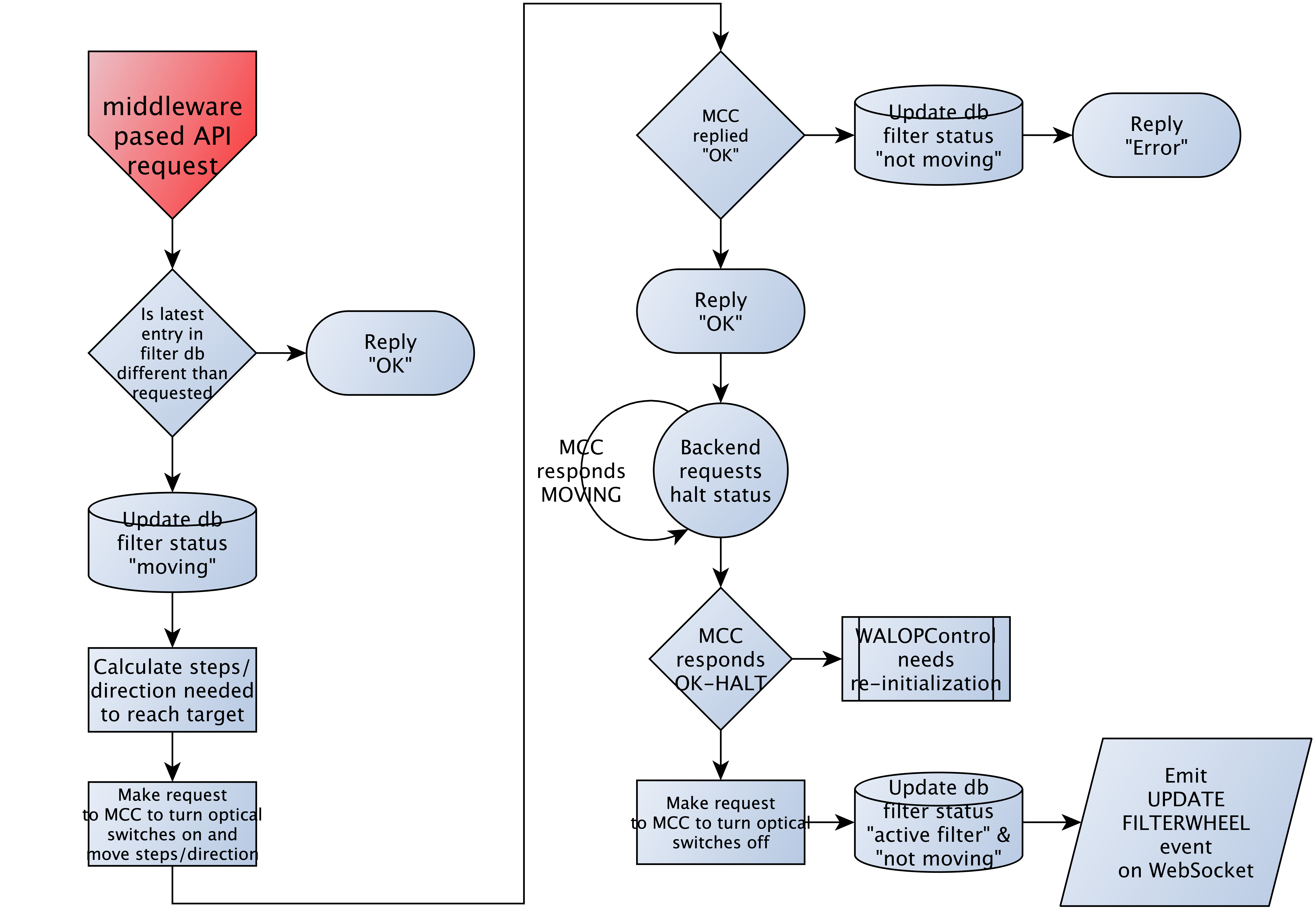}
\end{tabular}
\end{center}
\caption 
{ \label{fig:filtmove}
The handling of the "set filter" API call to the WALOPControl-backend.}
\end{figure}

The ``set filter" API call, is handled by first comparing the currently active filter (the latest entry in the filter-wheel database) with the requested one. If they are different, the difference in the motor position between the active and requested filter is calculated in steps. The database is updated with the latest filter-wheel update (same active filter as before the request, but now the filter-wheel is shown as moving). The user is notified that the filter-wheel status has changed by a WebSocket event emission. The MCC is alerted to turn on the optical switches and the required movement is requested to the MCC that handles the filter-wheel motor and a successful reply is sent to the user. The MCC is continuously polled for the motion status and upon non-erroneous halt, the updated filter status (active filter the requested and motion halted) is pushed to the database. The MCC is now requested to turn off the optical switches. Finally, the user is notified via the WebSocket event emission that the filter status has been updated once again and they can pull the latest status from the ``filter status" API call. The flowchart of the handling of the ``set filter" call is shown in Figure \ref{fig:filtmove}.

\subsubsection{Control of rotary motors}\label{subsubsec:rotary_motor}
All rotary motors also have a common control scheme. That means that the API endpoint is different for setting the calibration HWP rotation status, and the calibration polarizer rotation status, but the logic behind it is the same. We will give below an example of a request to set the HWP status. The linear part (moving the HWP in/out of the beam) is handled as Section \ref{subsubsec:linear_motor}.

For the rotary part, the API controller checks whether a change to the current status of the HWP is required. If no, it replies with \verb|200 - OK|. If yes, it creates a copy of the latest document of the HWP status collection in the database setting it as moving. Depending on whether the user asked for a continuous motion of the HWP or not, the controller will follow one of the following procedures:
\begin{itemize}
    \item\textbf{If continuous motion was requested:} WALOPControl-backend requests from the MCC to start (or stop) continuous motion of the HWP motor in the direction and speed requested by the user. Upon successful completion of the request, WALOPControl-backend replies to the user with \verb|200 - OK|, updates the HWP status collection of the database to reflect the change, and emits a HWP status update event on the WebSocket.
    \item\textbf{If finite motion was requested:} The database is updated with the latest HWP status update (same angle as before the request, but now the HWP is shown as moving). The user is notified that the HWP status has changed by a WebSocket event emission. WALOPControl-backend calculates the required amount of steps for the angle change and requests from the MCC to perform the movement, having first turned the optical switches on. The user receives an \verb|200 - OK| reply. The MCC is continuously polled for the motion status and upon non-erroneous halt, the database is updated to reflect the angle change and halt and a HWP status update event is emitted on the WebSocket.
\end{itemize}

\begin{figure}[!ht]
\begin{center}
\begin{tabular}{c}
\includegraphics[width=0.95\textwidth]{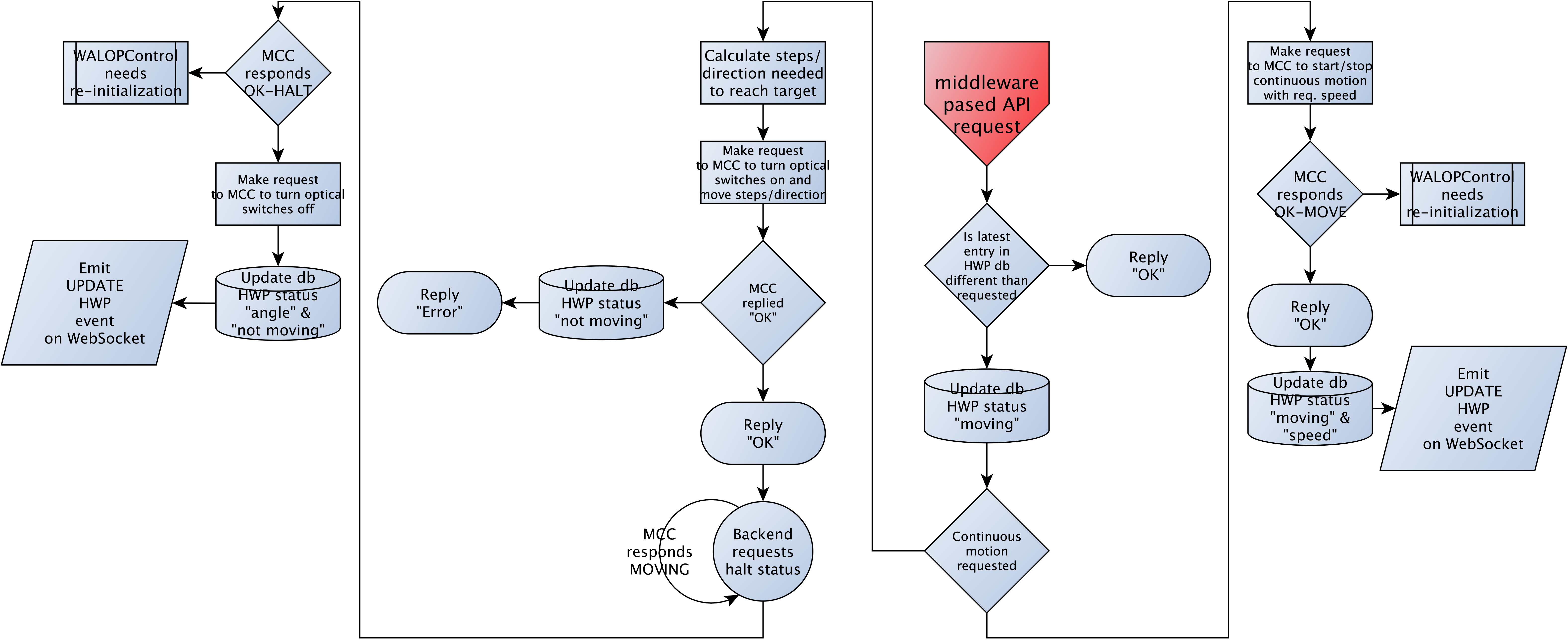}
\end{tabular}
\end{center}
\caption 
{ \label{fig:hwpmove}
The "set HWP" controller of the WALOPControl-backend. The red tile is the entrypoint of the controller, while the blue ones are the steps that are executed in order to handle the request.}
\end{figure}

The flowchart of the handling of the ``set hwp" call is shown in Figure \ref{fig:hwpmove}.

\subsubsection{Telescope control}\label{subsubsec:telescope}

The telescope API controller is very simple. The ``get status" API controller polls the database for the latest document in the telescope status collection (Section \ref{sect:db}), and responds to the user with \verb|200 - OK| and the contents of said document. These documents are created by the background job that periodically polls the TCS, started during the boot of the WALOPControl-backend (Section \ref{sect:tcs_init}).

The ``set next slew target" API controller simply pushes a document to the telescope status database, updating the next target to the requested, responds with \verb|200 - OK| and emits a telescope status update request.

The ``slew to next target" API controller requests from the TCS to start slewing to the next target, and upon successful request responds to the user with \verb|200 - OK|. It also increases the polling frequency of the TCS status to once every 5 seconds, until 30 seconds after slewing is completed or interrupted.

The ``stop slewing" API controller requests from the TCS to stop slewing, and upon successful request responds to the user with \verb|200 - OK|.

\subsubsection{Weather data exchange}\label{subsubsec:weather}

The weather API offers a single route. The ``get status" API controller polls the database for the latest document in the weather status collection (Section \ref{sect:db}), and responds to the user with \verb|200 - OK| and the contents of said document. These documents are created by the background job that periodically polls the weather station, started during the boot of the WALOPControl-backend (Section \ref{sect:wea_init}).

\subsubsection{CCD Control}\label{subsubsec:camera}

The CCD API offers 3 routes: to get the status of the CCDs, to request an exposure, and to abort a running exposure. The ``get status" API controller polls the database for the latest document in the CCD status collection (Section \ref{sect:db}), and responds to the user with \verb|200 - OK| and the contents of said document.

The ``expose" API controller adds a document to the CCD database for the observation to come, marking it as underway. \todo{It causes} WALOPControl-camera to start the observation, and upon successful request responds to the user with \verb|200 - OK|. WALOPControl-camera handles all CCD operations and timing (with a precision of $15\mu{}s$) up to the creation of the FITS file. It does this by setting up all IDSAC controllers in series, and after all are set up, it gives the expose command to the IDSAC from 4 cores of the CPU of its host machine simultaneously, one for each IDSAC. When WALOPControl-camera has finished saving the fits file (in the same hard drive that the backend is using), it sends a special API request to WALOPControl-backend notifying it of the change. WALOPControl-backend then updates the latest document in the CCD collection of the database (reflecting the successful exposure), creates a downscaled copy \todo{(binned to a resolution of 512x512 pixels)} of the fits file (for quick reference), and emits a CCD update event in the WebSocket. \todo{The binned copy is used to reduce the bandwidth usage between the backend and the frontend, since the full resolution images are not needed for most of the operations in the frontend. The binned copy is also used to display the latest image in the frontend, while the full resolution image is used for further analysis. Ad hoc analysis of the full resolution image can be done by downloading the full resolution image from the backend server.}

The ``stop exposure" API controller requests from WALOPControl-camera to stop the active exposures, and upon successful request responds to the user with \verb|200 - OK|. It then updates the latest document in the CCD collection of the database, reflecting the failed exposure, and emits a CCD update event in the WebSocket.

\section{WALOPCamera}\label{sect:camera}

The WALOPControl-camera software is a separate piece of software that is responsible for controlling the CCDs of the WALOP instrument. It is written in C and uses the IDSAC API library\cite{idsacapi} to communicate with the IDSAC cards. The software is installed on a small-form-factor PC (SFFPC) within the WALOP instruments. The software is responsible for receiving commands (related to the CCD operations) from WALOPControl-backend over TCP/IP and controlling the CCDs (according to those commands) via the IDSAC cards. The commands available to the WALOPCamera software user are listed in Appendix \ref{sect:ap_walopcamera}.

\subsection{Setting up the exposure}\label{subsec:exposure}

All exposures are set up in a persistent manner by the use of the \verb|TEMP|, \verb|GAIN| and \verb|FREQ| commands of WALOPCamera, setting the temperature, gain and readout frequency of upcoming observations respectivelly. In theory they need only be used once at the beginning of the night, since their settings persist through observations. Nevertheless, WALOPControl-backend will use these commands every time a new observation block is requested. Gain and readout frequency settings are set by moving the voltage and clock description files to the IDSAC-API library, while CCD temperature settings are passed to the temperature control card of the CCD.\\

Additionally, even though the \verb|EXPO| and \verb|MULT| commands persist the exposure time and multiples settings of the exposure block, WALOPControl-backend will use them every time a new observation block is requested. These settings are used to create the "exposure description" file, used by the IDSAC-API library in order to control the CCDs.\\

Finally, the \verb|SUBF| and \verb|BINN| commands are used to set the subframe and binning of the CCDs, respectively. These settings are not persistent and are set every time a new observation block is requested. These settings are also used in the creation of the "exposure description" file.

\subsection{Exposing the CCDs}\label{subsec:expose}

After the \verb|EXPO| command is issued, WALOPCamera first prepares the "exposure description" file and moves it to the IDSAC-API library folder. The IDSAC-API is then issued commands to clear the CCDs, reset the clocks and voltages, and start the exposure.\\

This procedure, up to the exposure start happens in serial for each of the 4 CCDs, yet the exposure command is issued to the API in parallel for all 4 CCDs. The result is a very accurate timing of the 4 exposure initiations (since it only depends on the SFFPC's CPU timing and no I/O operations), with a measured precision of $15\mu{}s$.\\

While the CCDs are exposing, the WALOPCamera software refuses any commands, other than \verb|STOP| and \verb|STAT|. The \verb|STAT| command is used to check the status of the CCDs, and is used by WALOPControl-backend in regular intervals to check if the exposure has finished.\\

When the exposure finishes, the WALOPCamera software saves the readout to a FITS file in the SFFPC's hard drive, using the header values provided by the \verb|HEAD| command. The readout is saved in a 3-axis ($n\times{}m\times{}4$, where n and m are determined by the subframe and binning), 16-bit, single-extension FITS file. WALOPCamera then sends an  API request to WALOPControl-backend notifying it of the change, and starts accepting commands again.\\

WALOPControl-backend will then update the latest document in the exposures collection of the database (reflecting the successful exposure), pulls (over FTP) the FITS file, creates a downscaled copy of the image (for quick reference), and emits a CCD update event in the WebSocket.

\section{WALOPControl Frontend}\label{sect:front}

The frontend of WALOPControl is written in the react.js\cite{react} framework, using the redux.js\cite{redux} framework for session management and the material UI component library\cite{material}. The choice of frameworks was made to ensure the fastest possible experience in the frontend, while optimizing user experience. The frontend is designed to be responsive, and to work on all modern browsers and devices. It provides a modern GUI to the user, consuming all the API and WebSocket offered by the backend. Appendix \ref{sect:ap_gui} provides a ``car-manual" demonstration of the GUI. A constantly updated video library of GUI demonstrations and combined backend-frontend tests (Section \ref{sect:test}) is available on \href{https://www.youtube.com/playlist?list=PLtVadXKiv58N8A3n9caW38qP1YAi-CuIz}{YouTube}.

\todo{The frontend is hosted on the same server as the backend, but is accessible outside the local network of the observatory, over a given domain. The frontend browser recipe is ``last 5 chrome versions, last 5 firefox versions, last 5 safari versions", which means that on build, react will make the frontend compatible with the latest 5 versions of these web browsers. At the time of writting the supported browsers are Firefox 134.0-138.0, Chrome 131-135 and Safari 15.4-17.6. Newer updates of WALOPControl will naturally support later versions of web browsers. Additionally, the frontend is responsive to all screen sizes and can be used by touch as well as mouse-controlled devices.}

The server is an nginx\cite{nginx} 1.16.0 and is set up to serve the frontend on the root of the domain, and the backend on the ``/api" path. The frontend is served as a static build, created by the react.js build command. The frontend is served over HTTPS, with a valid SSL certificate, while the backend is protected with a CORS (Cross-Origin Resource Sharing) filter, allowing only requests via the frontend or from other user within the respective observatory's network.

\section{WALOPControl Database}\label{sect:db}

WALOPControl's database is a MongoDB\cite{mongo} instance hosted on the Atlas cloud\cite{atlas} platform. By utilizing the capabilities of a cloud-hosted NoSQL database such as MongoDB, we provide users with nearly zero downtime and full compatibility with contemporary programming languages. 

\todo{The choice of a NoSQL database, specifically MongoDB, was made to accommodate the dynamic nature of the WALOPControl software. NoSQL databases are designed to handle unstructured data and can easily adapt to changes in the data model without requiring complex migrations. This flexibility is crucial for WALOPControl, as it allows for rapid development and iteration without the constraints of a rigid schema.}

\todo{One tradeoff of using a NoSQL database is the potential for data redundancy and inconsistency, as there are no enforced relationships between collections. However, this is mitigated in WALOPControl by carefully designing the collections and ensuring that data integrity is maintained through application logic.}

\todo{Another tradeoff is the lack of complex querying capabilities compared to traditional SQL databases. However, MongoDB provides powerful indexing and aggregation features that allow for efficient querying of large datasets, making it suitable for the needs of WALOPControl.}

In MongoDB, tables are referred to as \textit{collections}, and table entries are known as \textit{documents}.

The collections in the WALOPControl database are crafted to mirror the structure of both the backend and frontend while optimizing efficiency for read and write operations. The database is largely normalized, except for the logs collection, which is denormalized to enhance performance. Security is a priority, with user passwords hashed using SHA-256\cite{sha256_standard} before storage. To ensure speed, indexes are applied to all frequently queried fields. Additionally, the database is built for scalability, allowing for the addition of more servers to the cluster as needed.

The database structure is presented in list form in Appendix \ref{sect:ap_db_list} and in schematic form in Appendix \ref{sect:ap_db}.

\section{Continuous Integration and Continuous Deployment}\label{sect:cicd}

In order to ensure the quality of the WALOPControl software, we have set up a Continuous Integration and Continuous Deployment (CI/CD) pipeline. The software is hosted in a gitea\cite{gitea} repository, and as such the CI/CD pipeline written in Gitea's workflow format. Upon a change being made to the software code, and its repository being pushed to the institute's Gitea infrastructure, Gitea runners assume the CI/CD pipeline and all DevOps (Software Development Operations\cite{devops}), with an exception to end-to-end testing. DevOps in the spec of WALOPControl include:
\begin{enumerate}
    \item Unit Testing
    \item Integration Testing
    \item End-to-end testing
    \item Deployment to production
\end{enumerate}

\subsection{Unit Testing}\label{subsec:unit}
Unit testing is conducted whenever code is pushed to any branch of the repository \todo{(with support for testing in parallel and in isolation between branches, if code is pushed to multiple branches at a time)}. Automated tests are executed \todo{on the bare code (except in the case of the master branch - Section \ref{subsec:deploy})} to ensure that each middleware and controller in the backend functions correctly. Predefined test cases for each middleware and controller are evaluated independently in a controlled environment. The actual results are compared with the expected outcomes, and if any discrepancies are found, the subsequent DevOps processes are halted.

\subsubsection{Unit Testing of the Filter-wheel controller}\label{subsubsec:unit_filt}
The filter-wheel controller is tested by sending commands to a mock MCC and checking if the response is as expected.

\subsubsection{Unit Testing of the Calibration HWP controller}\label{subsubsec:unit_hwp}
The calibration HWP controller is tested by sending commands to a mock MCC and checking if the response is as expected.

\subsubsection{Unit Testing of the Calibration Polarizer controller}\label{subsubsec:unit_pol}
The calibration polarizer controller is tested by sending commands to a mock MCC and checking if the response is as expected.

\subsubsection{Unit Testing of the Focuser controller}\label{subsubsec:unit_foc}
The focuser controller is tested by sending commands to a mock MCC and checking if the response is as expected.

\subsubsection{Unit Testing of the Guider controller}\label{subsubsec:unit_gui}
The guider controller is tested by sending commands to a mock MCC and checking if the response is as expected.

\subsubsection{Unit Testing of the Telescope controller}\label{subsubsec:unit_tel}
The telescope controller is tested by sending commands to a mock TCS and checking if the response is as expected.

\subsubsection{Unit Testing of the Weather controller}\label{subsubsec:unit_wea}
The weather controller is tested by sending commands to a mock weather station and checking if the response is as expected.

\subsubsection{Unit Testing of the CCD controller}\label{subsubsec:unit_cam}
The CCD controller is tested by sending commands to an IDSAC controller not connected to any CCD. The timing of the expose commands to each CCD is measured with an acceptance cutoff of 20ms. At this stage the connection bandwidth between the backend and the IDSAC is also tested with a 30 second exposure-to-fits file cutoff.

\subsection{Integration Testing}\label{subsec:integr}
Integration testing is triggered each time code is pushed to any branch of the repository \todo{(with support for testing in parallel and in isolation between branches, if code is pushed to multiple branches at a time)}. Automated tests are run \todo{on the bare code (except in the case of the master branch - Section \ref{subsec:deploy})} to ensure that middlewares and controllers collaborate as intended. Predefined test cases for each API route are assessed independently in a controlled environment. The actual results are matched against the expected outcomes, and if there are any discrepancies, subsequent DevOps steps are blocked.

\subsubsection{Integration Testing of the Filter-wheel Route}\label{subsubsec:integ_filt}
The filter-wheel route is tested by sending a request to the ``get status" API endpoint, and checking if the response is \verb|200 - OK| and the contents of the response are the same as the latest document in the filter-wheel status collection of the database. The controller is also tested by sending a request to the ``set filter" API endpoint, and checking if the response is \verb|200 - OK| when an existing filter is requested and denied when a non-existing filter is requested.

\subsubsection{Integration Testing of the Calibration HWP Route}\label{subsubsec:integ_hwp}
The calibration HWP route is tested by sending a request to the ``get status" API endpoint, and checking if the response is \verb|200 - OK| and the contents of the response are the same as the latest document in the calibration HWP status collection of the database. The controller is also tested by sending a request to the ``set position" API endpoint, and checking if the response is \verb|200 - OK| when a valid position is requested and denied when an invalid position is requested.

\subsubsection{Integration Testing of the Calibration Polarizer Route}\label{subsubsec:integ_pol}
The calibration polarizer route is tested by sending a request to the ``get status" API endpoint, and checking if the response is \verb|200 - OK| and the contents of the response are the same as the latest document in the calibration polarizer status collection of the database. The controller is also tested by sending a request to the ``set position" API endpoint, and checking if the response is \verb|200 - OK| when a valid position is requested and denied when an invalid position is requested.

\subsubsection{Integration Testing of the Focuser Route}\label{subsubsec:integ_foc}
The focuser route is tested by sending a request to the ``get status" API endpoint, and checking if the response is \verb|200 - OK| and the contents of the response are the same as the latest document in the focuser status collection of the database. The controller is also tested by sending a request to the ``set focus position" API endpoint for each focuser, and checking if the response is \verb|200 - OK| when a valid position is requested and denied when an invalid position is requested.

\subsubsection{Integration Testing of the Guider Route}\label{subsubsec:integ_gui}
The guider route is tested by sending a request to the ``get status" API endpoint, and checking if the response is \verb|200 - OK| and the contents of the response are the same as the latest document in the guider status collection of the database. The controller is also tested by sending a request to the ``change positioning" API endpoint, and checking if the response is \verb|200 - OK| when a valid position is requested and denied when an invalid position is requested.

\subsubsection{Integration Testing of the Telescope Route}\label{subsubsec:integ_tel}
The telescope route is tested by sending a request to the ``get status" API endpoint, and checking if the response is \verb|200 - OK| and the contents of the response are the same as the latest document in the telescope status collection of the database. The controller is also tested by sending a request to the ``set next slew target" API endpoint, and checking if the response is \verb|200 - OK|. The test includes providing a mocked telescope software instance with the proper underlying command. The ``slew to next target" and ``stop slewing" API endpoints are tested in a similar manner, checking if the response is \verb|200 - OK| using a mocked telescope software instance.

\subsubsection{Integration Testing of the Weather Route}\label{subsubsec:integ_wea}
The weather route is tested by sending a request to the ``get status" API endpoint, and checking if the response is \verb|200 - OK| and the contents of the response are the same as the latest document in the weather status collection of the database.

\subsubsection{Integration Testing of the CCD Route}\label{subsubsec:integ_cam}
The CCD route is tested by sending a request to the ``get status" API endpoint, and checking if the response is \verb|200 - OK| and the contents of the response are the same as the latest document in the CCD status collection of the database. The controller is also tested by sending a request to the ``expose" API endpoint, and checking if the response is \verb|200 - OK| and a proper mock fits file is saved. The controller is also tested by sending a request to the ``stop exposure" API endpoint, and checking if the response is \verb|200 - OK|.

\subsection{Deployment to production}\label{subsec:deploy}
Upon successful administration of unit and integration tests, after the code \todo{is merged} to the master branch of the git repository, the lead developer has the right to allow the pipeline to proceed to deployment. \todo{They} only do so, after ensuring \todo{themselves} that all tests did pass (not trusting the pipeline output)\todo{, and solving any merge conflicts that may arise}. The code is compiled into a Docker\cite{docker} image, the image is pushed to the gitea repository image registry, and the server machine is notified (via a rest API server daemon) to download the updated image and start the container that hosts it. \todo{The same unit and integration tests that were ran by the CI/CD pipeline on the bare code, are also ran by it on the Docker image before it is pushed to the registry. This happens by having written separate testing CI/CD pipelines for the master branch. Having tests ran by Docker images has proved to be slower, thus we have reserved them only for the master branch which will ultimately be deployed to production. Nevertheless, testing on the images is important, since it ensures that the code will run as expected on the server, and that no environment-specific issues will arise.}

We chose Docker since its unparalleled universality allows us the flexibility of host machines for WALOPControl. The downtime of the server in case of an update, using Docker and the automatic CI/CD pipeline is less than 30 seconds and the update release time (from merging to the master branch to initiating server update) is less than 15 minutes, including unit testing, integration testing, image compilation, and image upload, with most time spent on unit and integration testing \todo{(maximum recorded time for all unit tests in master was 6 minutes - same for all integration tests)}.

\subsection{Versioning}\label{subsec:versioning}
\todo{WALOPControl uses semantic versioning (SemVer\cite{semantic}) to track changes in the software. This is a versioning scheme that uses a three-part version number: MAJOR.MINOR.PATCH. MAJOR version changes indicate incompatible API changes, MINOR version changes indicate backward-compatible functionality, and PATCH version changes indicate backward-compatible bug fixes.}

\todo{The version number is incremented according to the following rules:
\begin{itemize}
    \item If a new feature is added, the MINOR version is incremented.
    \item If a bug is fixed, the PATCH version is incremented.
    \item If a breaking change is introduced, the MAJOR version is incremented and the MINOR and PATCH versions are reset to 0.
\end{itemize}}

\todo{The version number is stored in the "package.json"\cite{isaac_z_schlueter_npm_2010} file of the WALOPControl-backend and WALOPControl-frontend, and is also included in the Docker image tags. The version number is also displayed in the GUI of the WALOPControl-frontend.}

\subsubsection{Versioning of the technologies used}\label{subsubsec:versioning_tech}
\todo{At the time of writing, WALOPControl (version 4.3.8) uses the following core technologies, with the respective versions:
\begin{itemize}
    \item Node.js version 22.11.0
    \item Express.js version 4.18.2
    \item MongoDB version 8.0.10
    \item React.js version 18.2.0
    \item Redux.js version 8.0.5
    \item Material UI version 5.12.3
    \item nginx version 1.16.0
\end{itemize}}

\todo{The versions of the technologies used are frozen upon compilation of the Docker image, and are not updated until a security update to them enforces it. This ensures that the software is always compatible with the technologies used, and that no breaking changes are introduced without proper testing. If any of the technologies used is updated, the version number of WALOPControl is incremented accordingly, and the software is tested to ensure that it works correctly with the new version of the technology, as specified in Section \ref{sect:test}.}

\section{End-To-End Testing}\label{sect:test}
End-to-end testing refers to tests performed at the lab, using a developer-only version of the code. This is conducted by a minimum of 2 members of the PASIPHAE collaboration, or the Instrumentation Laboratory of the Inter University Centre for Astronomy and Astrophysics (IUCAA). It involves testing the software end-to-end, using the GUI to control the instrument, and checking the result of predefined sequences of inputs to the software. 1 of the testers is using the GUI to control the instrument, and the other is checking the response of the instrument. The instrument response is measured using micrometers (for the motors) and checking FITS images produced by the CCDs for defects. A library of such tests is available on \href{https://www.youtube.com/playlist?list=PLtVadXKiv58N8A3n9caW38qP1YAi-CuIz}{YouTube}.

\todo{The following conditions will trigger an end-to-end test:
\begin{enumerate}
    \item A new major or minor version of the software is ready for release.
    \item Any hardware change is made to the instrument.
    \item Any library/technology used by the software is updated to a new major or minor version.
    \item Any change is made to the database structure, even if condition 1 is not met.
    \item Any change is made to the WALOPCamera software (since it does not track the version changes of WALOPControl).
\end{enumerate}}

\subsection{Precision testing}\label{subsec:precision}
\todo{As part of a non-standard ent-to-end testing, the precision in which the software can command the instrument is tested. This is done using an automated ``schedule'', uploaded to the frontend in a similar manner to an observation schedule. The schedule consists of a series of commands to the instrument, moving each component sequentially between given positions, and checking the response of the instrument. The precision is measured by the difference between the commanded position and the actual position of the component, as measured by a micrometer. The precision is expected to be within the limit of $1\mu{}m$ for all measurements, as detailed in Section \ref{sect:reqs}.}

\todo{This non-standard end-to-end test is conducted with the same conditions as the standard end-to-end test, with the addition of a micrometer to measure the position of the components. Triggers of this test are the addition of a new component to the instrument, a change to the hardware of an existing component, or a change in the software that may affect the precision of the instrument.}

\subsection{Fault testing}\label{subsec:fault}
\todo{Another non-standard end-to-end testing is fault testing. Fault testing is conducted to ensure that the software can handle unexpected situations and errors gracefully. This includes testing the software's response to hardware failures, network issues, and other unexpected situations. To simulate these events we run otherwise standard end-to-end tests (by using automated ``schedules'' as described in Section \ref{subsec:precision}), but we introduce live faults in the system, such as disconnecting the hardware, network, or introducing mechanical faults (e.g. motor stops). The software is expected to handle these situations gracefully, without crashing or producing incorrect results. The software should also provide meaningful error messages to the user and notify the admins in case of non-automatic recovery.}

\todo{Triggers of this test are the addition of a new component to the instrument, or a change in the software that may affect the fault tolerance of the instrument. For example, the addition of a new admin notification method would trigger this test, to ensure that the software can handle the new notification system properly.}

\subsection{On-Sky Testing}\label{subsec:onsky}
\todo{Even though our extensive end-to-end testing in the lab has shown us that the software works as intended, we still intend to test it on-sky regularly. This is expected to be done in nights marked as ``maintenance nights'', using the same conditions as end-to-end testing but on real hardware instead of a lab setup. The tests can also include precision and fault testing, as described in Sections \ref{subsec:precision} and \ref{subsec:fault}. The on-sky testing is expected to be conducted at least once yearly to ensure the quality of services provided by the instrument and software.}

\section{Conclusions}\label{sect:concl}

WALOPControl is a modern, efficient, and secure software package that allows for the control of the WALOP instruments at the Skinakas Observatory and the South African Astronomical Observatory. It is designed to be user-friendly, with a modern - responsive GUI, and to be as efficient as possible, with a fast backend and a cloud-hosted database. It is also designed to be as secure as possible, with user authentication and authorization, and reliable thanks to its CI/CD pipeline and end-to-end testing. Maintainability is ensured, with a clean codebase and extensive documentation. It is designed to be extensible, with the ability to add new features and instruments as needed, as well as accept software extensions on a unified backend.

\appendix    

\section{WALOP-North Controllables}\label{sect:ap_north}
Table \ref{tab:walopn} presents the controllables and telematics of the WALOP-North instrument.

\begin{table}[ht]
\caption{Controllables and telematics of the WALOP-North instrument.} 
\label{tab:walopn}
\begin{center}       
\begin{tabular}{||||||c|c||||||} 
\hline\hline\hline
\hline\hline\hline
\rule[-1ex]{0pt}{3.5ex}  Controllable & Interface  \\
\hline\hline\hline
\rule[-1ex]{0pt}{3.5ex}  Calibration Polarizer Placement (linear motion through lead screw) & MCC - TCP/IP  \\
\hline
\rule[-1ex]{0pt}{3.5ex}  Calibration Polarizer Rotation (rotary motion through gear coupling) & MCC - TCP/IP  \\
\hline
\rule[-1ex]{0pt}{3.5ex}  Calibration HWP Placement (linear motion through lead screw) & MCC - TCP/IP  \\
\hline
\rule[-1ex]{0pt}{3.5ex}  Calibration HWP Rotation (rotary motion through gear coupling) & MCC - TCP/IP  \\
\hline
\rule[-1ex]{0pt}{3.5ex}  Filter-wheel Position (linear motion through lead screw) & MCC - TCP/IP  \\
\hline
\rule[-1ex]{0pt}{3.5ex}  CCD 1 (exp. time, temperature, gain, readout frequency, subframe, binning) & IDSAC - USB  \\
\hline
\rule[-1ex]{0pt}{3.5ex}  CCD 2 (exp. time, temperature, gain, readout frequency, subframe, binning) & IDSAC - USB \\
\hline
\rule[-1ex]{0pt}{3.5ex}  CCD 3 (exp. time, temperature, gain, readout frequency, subframe, binning) & IDSAC - USB \\
\hline
\rule[-1ex]{0pt}{3.5ex}  CCD 4 (exp. time, temperature, gain, readout frequency, subframe, binning) & IDSAC - USB \\
\hline
\rule[-1ex]{0pt}{3.5ex}  CCD 1 Focuser Position (linear motion through lead screw) & MCC - TCP/IP  \\
\hline
\rule[-1ex]{0pt}{3.5ex}  CCD 2 Focuser Position (linear motion through lead screw) & MCC - TCP/IP  \\
\hline
\rule[-1ex]{0pt}{3.5ex}  CCD 3 Focuser Position (linear motion through lead screw) & MCC - TCP/IP  \\
\hline
\rule[-1ex]{0pt}{3.5ex}  CCD 4 Focuser Position (linear motion through lead screw) & MCC - TCP/IP  \\
\hline
\rule[-1ex]{0pt}{3.5ex}  Guider Position around FoV (rotary motion through gear coupling) & MCC - TCP/IP  \\
\hline
\rule[-1ex]{0pt}{3.5ex}  1.3m Telescope Control (equatorial motion \& tracking) & TCP/IP  \\
\hline\hline\hline
\hline\hline\hline
\rule[-1ex]{0pt}{3.5ex}  Telematic & Interface  \\
\hline\hline\hline
\rule[-1ex]{0pt}{3.5ex}  1.3m Telescope Telematics & TCP/IP  \\
\hline
\rule[-1ex]{0pt}{3.5ex}  Weather at Skinakas Observatory & Alpaca  \\
\hline\hline\hline
\hline\hline\hline
\end{tabular}
\end{center}
\end{table} 

\section{WALOP-South Controllables}\label{sect:ap_south}
Table \ref{tab:walops} presents the controllables and telematics of the WALOP-South instrument.
\begin{table}[ht]
\caption{Controllables and telematics of the WALOP-South instrument.} 
\label{tab:walops}
\begin{center}       
\begin{tabular}{||||||c|c||||||} 
\hline\hline\hline
\hline\hline\hline
\rule[-1ex]{0pt}{3.5ex}  Controllable & Interface  \\
\hline\hline\hline
\rule[-1ex]{0pt}{3.5ex}  Calibration Polarizer Placement (linear motion through lead screw) & MCC - TCP/IP  \\
\hline
\rule[-1ex]{0pt}{3.5ex}  Calibration Polarizer Rotation (rotary motion through gear coupling) & MCC - TCP/IP  \\
\hline
\rule[-1ex]{0pt}{3.5ex}  Calibration HWP Placement (linear motion through lead screw) & MCC - TCP/IP  \\
\hline
\rule[-1ex]{0pt}{3.5ex}  Calibration HWP Rotation (rotary motion through gear coupling) & MCC - TCP/IP  \\
\hline
\rule[-1ex]{0pt}{3.5ex}  Filter-wheel Position (linear motion through lead screw) & MCC - TCP/IP  \\
\hline
\rule[-1ex]{0pt}{3.5ex}  CCD 1 (exp. time, temperature, gain, readout frequency, subframe, binning) & IDSAC - USB  \\
\hline
\rule[-1ex]{0pt}{3.5ex}  CCD 2 (exp. time, temperature, gain, readout frequency, subframe, binning) & IDSAC - USB  \\
\hline
\rule[-1ex]{0pt}{3.5ex}  CCD 3 (exp. time, temperature, gain, readout frequency, subframe, binning) & IDSAC - USB \\
\hline
\rule[-1ex]{0pt}{3.5ex}  CCD 4 (exp. time, temperature, gain, readout frequency, subframe, binning) & IDSAC - USB \\
\hline
\rule[-1ex]{0pt}{3.5ex}  CCD 1 Focuser Position (linear motion through lead screw) & MCC - TCP/IP  \\
\hline
\rule[-1ex]{0pt}{3.5ex}  CCD 2 Focuser Position (linear motion through lead screw) & MCC - TCP/IP  \\
\hline
\rule[-1ex]{0pt}{3.5ex}  CCD 3 Focuser Position (linear motion through lead screw) & MCC - TCP/IP  \\
\hline
\rule[-1ex]{0pt}{3.5ex}  CCD 4 Focuser Position (linear motion through lead screw) & MCC - TCP/IP  \\
\hline
\rule[-1ex]{0pt}{3.5ex}  X Guider Position around FoV (linear motion through lead screw) & MCC - TCP/IP  \\
\hline
\rule[-1ex]{0pt}{3.5ex}  Y Guider Position around FoV (linear motion through lead screw) & MCC - TCP/IP  \\
\hline
\rule[-1ex]{0pt}{3.5ex}  1m Telescope Control (equatorial motion \& tracking) & TCP/IP  \\
\hline\hline\hline
\hline\hline\hline
\rule[-1ex]{0pt}{3.5ex}  Telematic & Interface  \\
\hline\hline\hline
\rule[-1ex]{0pt}{3.5ex}  1m Telescope Telematics & TCP/IP  \\
\hline
\rule[-1ex]{0pt}{3.5ex}  Weather at Skinakas Observatory & HTTP API  \\
\hline\hline\hline
\hline\hline\hline
\end{tabular}
\end{center}
\end{table} 

\section{Motion Control Card TCP/IP Protocol}\label{sect:ap_mcc}
The MCCs used by the WALOP polarimeters respond to commands over TCP/IP. All commands have a fixed format, as follows: \texttt{\$<COMMAND>,<SERIAL>,<ARGUMENTS>,<CHECKSUM>}. \texttt{<COMMAND>} is the name of the command to be executed (refer to list below), \texttt{<SERIAL>} is the serial number of the motor to which the command refers (a 3-digit number followed by a dash and a 6-digit number), \texttt{<ARGUMENTS>} are arguments specific to each command (optional - refer to list below), and \texttt{<CHECKSUM>} is the integrity checksum (1 byte long - 8bit, 2s complement sum of the entire command phrase). The commands are case-sensitive.

The MCCs respond to commands with a fixed format, as follows: \texttt{<CODE>,<PARAMETERS>}. \texttt{<CODE>} is the response code: either \texttt{01} (if no error) or a number greater than one if an error was encountered. \texttt{<PARAMETERS>} are parameters specific to each command (optional - refer to list below). The responses are case-sensitive.

The list below presents the command names that can be sent to the Motion Control Card (MCC) over TCP/IP and their arguments. These refer to any kind of motor control, irrespective of the motor's coupling (linear or rotary), unless otherwise specified. The motor in question is identified by its serial number, as provided in the command phrase.

\begin{itemize}
    \item \textbf{HOME} - Homes the motor to a specified end-switch position. No arguments accepted. No parameters returned. Refers only to linear motors.
    \item \textbf{HOMA} - Homes the motor to a specified end-switch position. No arguments accepted. No parameters returned. Refers only to rotary motors.
    \item \textbf{GMST} - Request the motion status of the motor. No arguments accepted. Returned parameter is the motion status of the motor as: \texttt{HALT} (if halted) or \texttt{MOVE} (if moving).
    \item \textbf{SFIN} - Request finite motion of the motor. 2 arguments are needed: the first one is the direction of movement (\texttt{+} or \texttt{-}) and the second one is the movement length in $steps$. No parameters returned.
    \item \textbf{SCON} - Set-up continuous motion of the motor. 2 arguments are needed: the first one is the direction of movement (\texttt{+} or \texttt{-}) and the second one is the movement speed in $\frac{steps}{s}$. No parameters returned.
    \item \textbf{STRT} - Starts continuous motion of the motor as set-up by \texttt{SCON}. No arguments accepted. No parameters returned.
    \item \textbf{STOP} - Stops the motion of the motor (either continuous or finite). No arguments accepted. No parameters returned.
    \item \textbf{SNON} - Turns on the light switches on the path of the motor (they need to be on while the motor performs finite motion or homing - danger of runaway motors if the light switches are off). No arguments accepted. No parameters returned.
    \item \textbf{SNOF} - Turns off the light switches on the path of the motor (they need to be off while the instrument observes to avoid interference). No arguments accepted. No parameters returned.
\end{itemize}

\section{WALOPCamera Commands}\label{sect:ap_walopcamera}
Table \ref{tab:walopcam} presents the commands acceptable by the WALOPCamera software.
\begin{table}[ht]
\caption{Commands acceptable by the WALOPCamera software.}
\label{tab:walopcam}
\begin{center}
\begin{tabular}{||||||c|c||||||}
\hline\hline\hline
\hline\hline\hline
\rule[-1ex]{0pt}{3.5ex}  Command & Description  \\
\hline\hline\hline
\rule[-1ex]{0pt}{3.5ex}  \texttt{SETT} & Set the SFFPC time  \\
\hline
\rule[-1ex]{0pt}{3.5ex}  \texttt{EXPO} & Set the exposure time of the exposure block - persists observations  \\
\hline
\rule[-1ex]{0pt}{3.5ex}  \texttt{MULT} & Set the number of exposures for the exposure block - persists observations  \\
\hline
\rule[-1ex]{0pt}{3.5ex}  \texttt{HEAD} & Set a header value for the fits file - persists observations  \\
\hline
\rule[-1ex]{0pt}{3.5ex}  \texttt{TEMP} & Set the temperature of the CCDs - persists observations  \\
\hline
\rule[-1ex]{0pt}{3.5ex}  \texttt{GAIN} & Set the gain of the CCDs - persists observations  \\
\hline
\rule[-1ex]{0pt}{3.5ex}  \texttt{FREQ} & Set the readout frequency of the CCDs - persists observations  \\
\hline
\rule[-1ex]{0pt}{3.5ex}  \texttt{SUBF} & Set the pixel subframe - resets after 1 exposure block  \\
\hline
\rule[-1ex]{0pt}{3.5ex}  \texttt{BINN} & Set the pixel binning - resets after 1 exposure block  \\
\hline
\rule[-1ex]{0pt}{3.5ex}  \texttt{STRT} & Start the exposure block  \\
\hline
\rule[-1ex]{0pt}{3.5ex}  \texttt{STOP} & Stop the exposure  \\
\hline
\rule[-1ex]{0pt}{3.5ex}  \texttt{STAT} & Get the status of the CCDs  \\
\hline\hline\hline
\hline\hline\hline
\end{tabular}
\end{center}
\end{table}

\section {GUI Demonstration}\label{sect:ap_gui}

\begin{figure}[!ht]
\begin{center}
\begin{tabular}{c}
\includegraphics[width=0.8\textwidth]{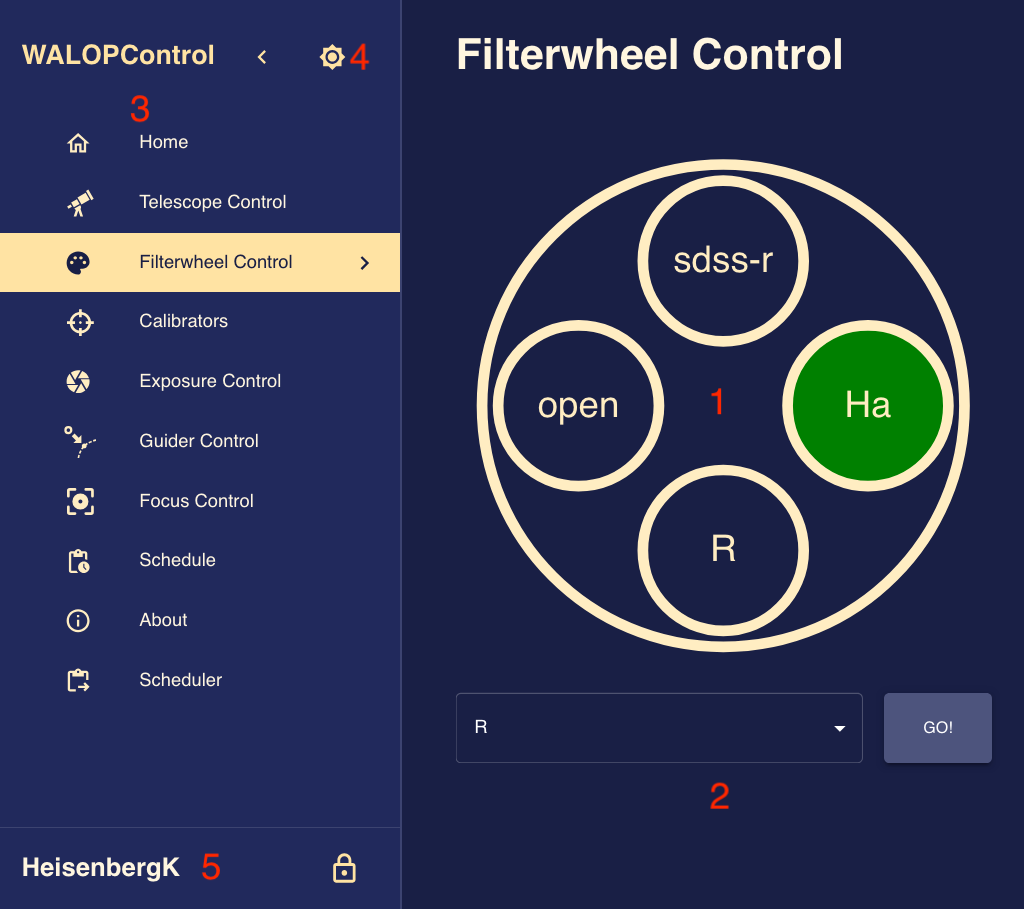}
\end{tabular}
\end{center}
\caption 
{ \label{fig:gui_fw}
The filterwheel control page of the WALOPControl GUI. 1: active filter display, 2: filter selection control, 3: sidebar common to all GUI pages (to select the active control page), 4: light/dark mode switch, 5: logged-in user - link to user settings. }
\end{figure}

\begin{figure}[!ht]
\begin{center}
\begin{tabular}{c}
\includegraphics[width=0.8\textwidth]{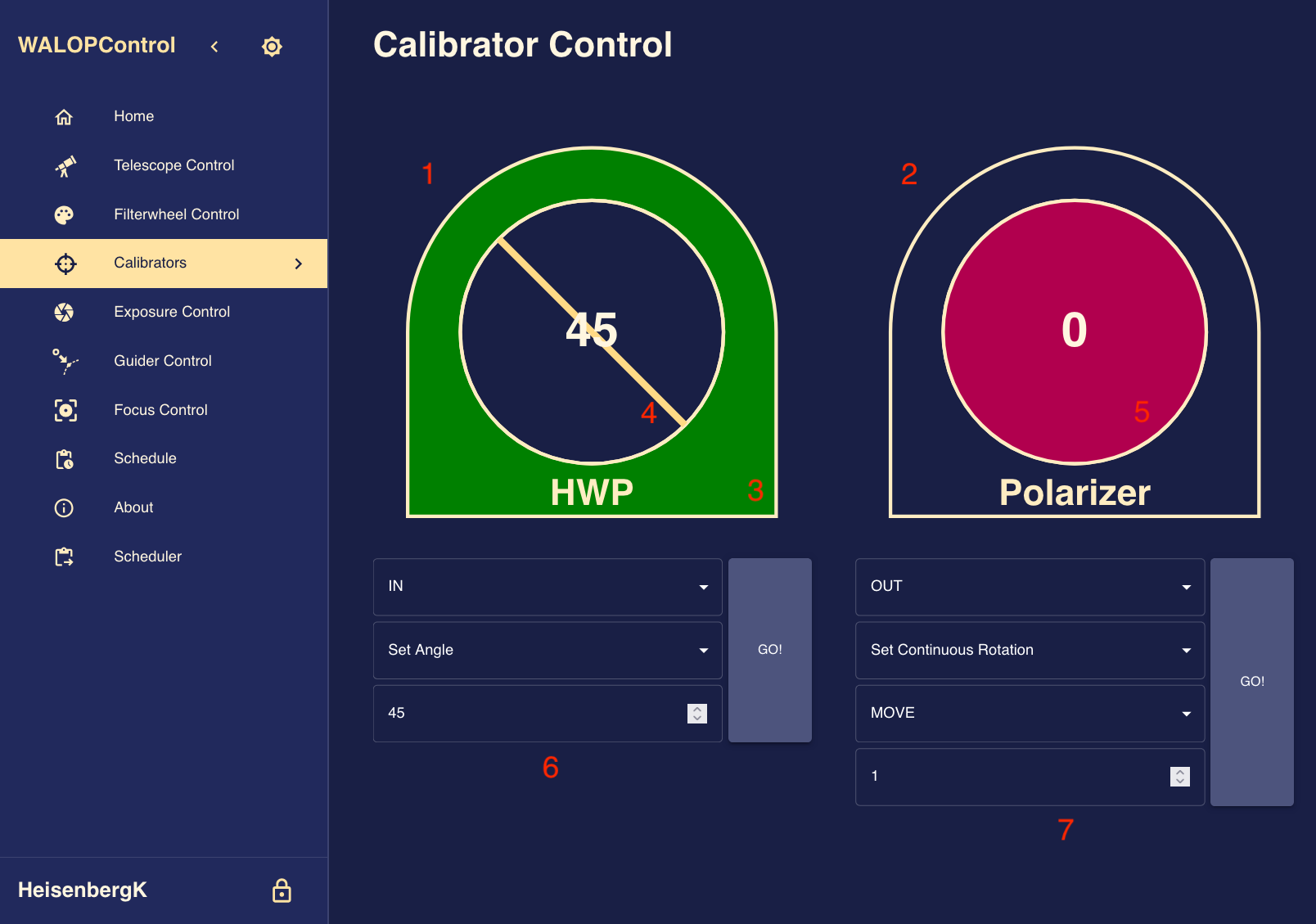}
\end{tabular}
\end{center}
\caption 
{ \label{fig:gui_cal}
The calibrators control page of the WALOPControl GUI. 1: Calibration HWP display, 2: Calibration Polarizer display, 3: either outer panel (of HWP or Polarizer) turns green when the controllable is in the optical path of the intrument, 4: when the controllable is in angle-select mode, the angle is diplayed as a line-and-number combo, 5: when the controllable is in continuous rotation mode, the the inner circular disk flashes red-greed indicating rotation and the angle display is unavailable, 6: Calibration HWP movement control, 7: Calibration Polarizer movemnt control. }
\end{figure}

\begin{figure}[!ht]
\begin{center}
\begin{tabular}{c}
\includegraphics[width=0.8\textwidth]{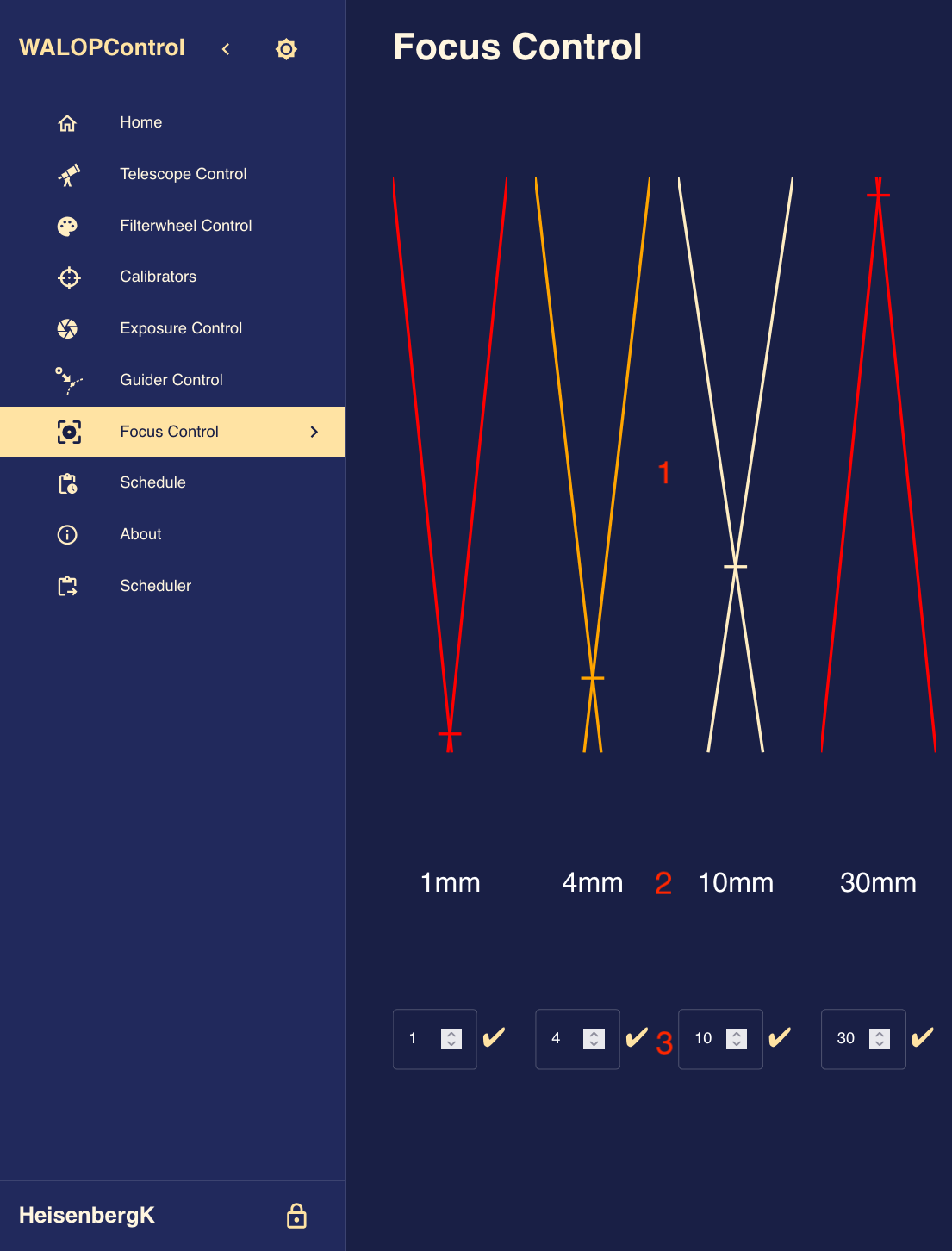}
\end{tabular}
\end{center}
\caption 
{ \label{fig:gui_focus}
The focuser control page of the WALOPControl GUI. 1: 4-focuser display, where red means very close to either limit, orange means approaching either limit, beige means no limit proximity 2: 4-focuser positions 3: 4-focuser controls. }
\end{figure}

\begin{figure}[!ht]
\begin{center}
\begin{tabular}{c}
\includegraphics[width=0.95\textwidth]{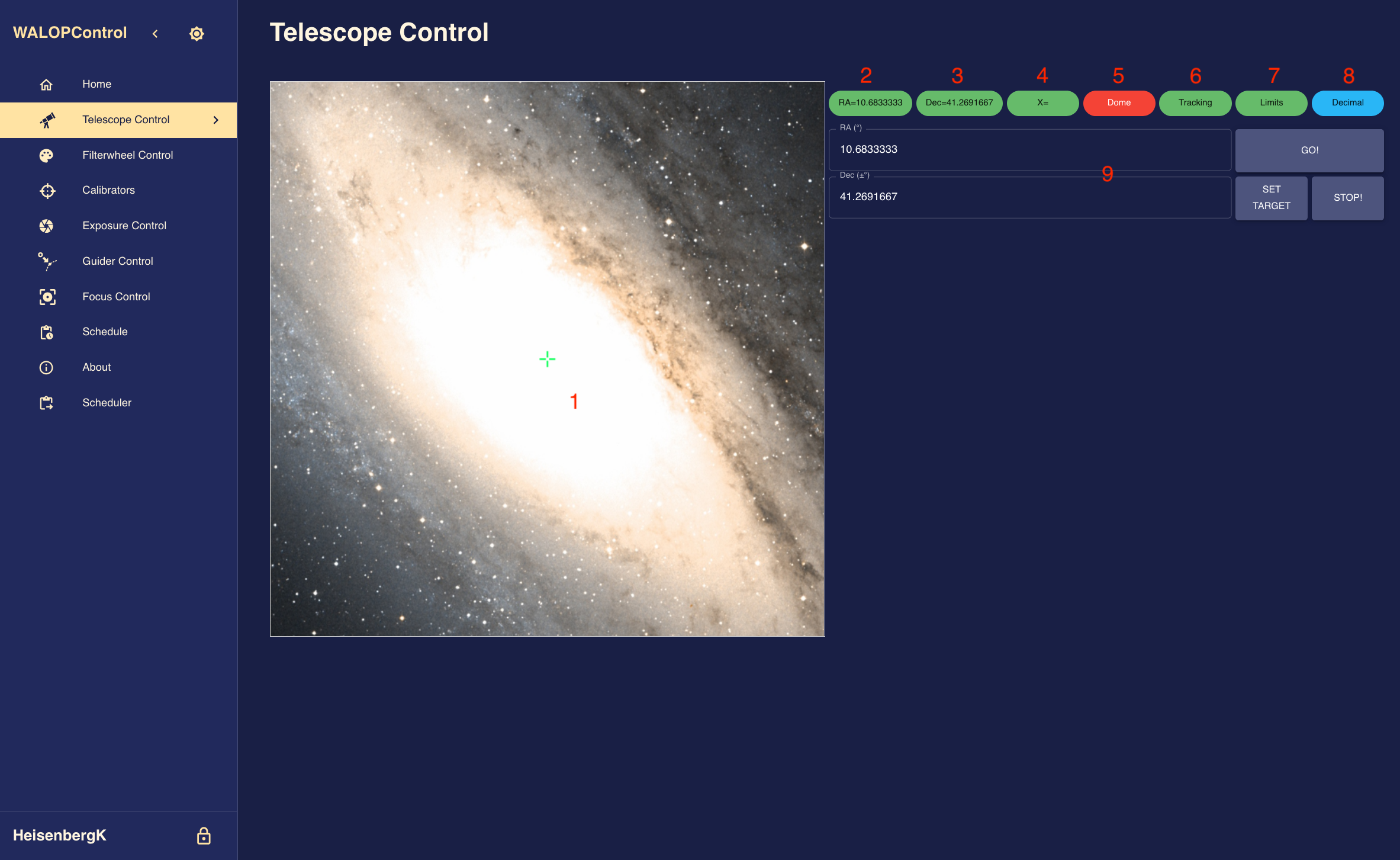}
\end{tabular}
\end{center}
\caption 
{ \label{fig:gui_telescope}
The telescope control page of the WALOPControl GUI. 1: current telescope pointing, scaled to the WALOP FoV, retrieved from Aladin \cite{aladin} , 2: current pointing R.A. of the telescope, 3: current pointing Dec. of the telescope, 4: current pointing Airmass of the telescope if $>3$, 5: red if dome shutter is closed or dome not tracking, green otherwise, 6: green if telescope is tracking, red otherwise, 7: green if no telescope limit proximity triggered, orange if telescope approaching limit, red if telescope at limit, 8: toggle telescope pointing control between desimal and sexagessimal mode, 9: telescope pointing control. }
\end{figure}

\begin{figure}[!ht]
\begin{center}
\begin{tabular}{c}
\includegraphics[width=0.95\textwidth]{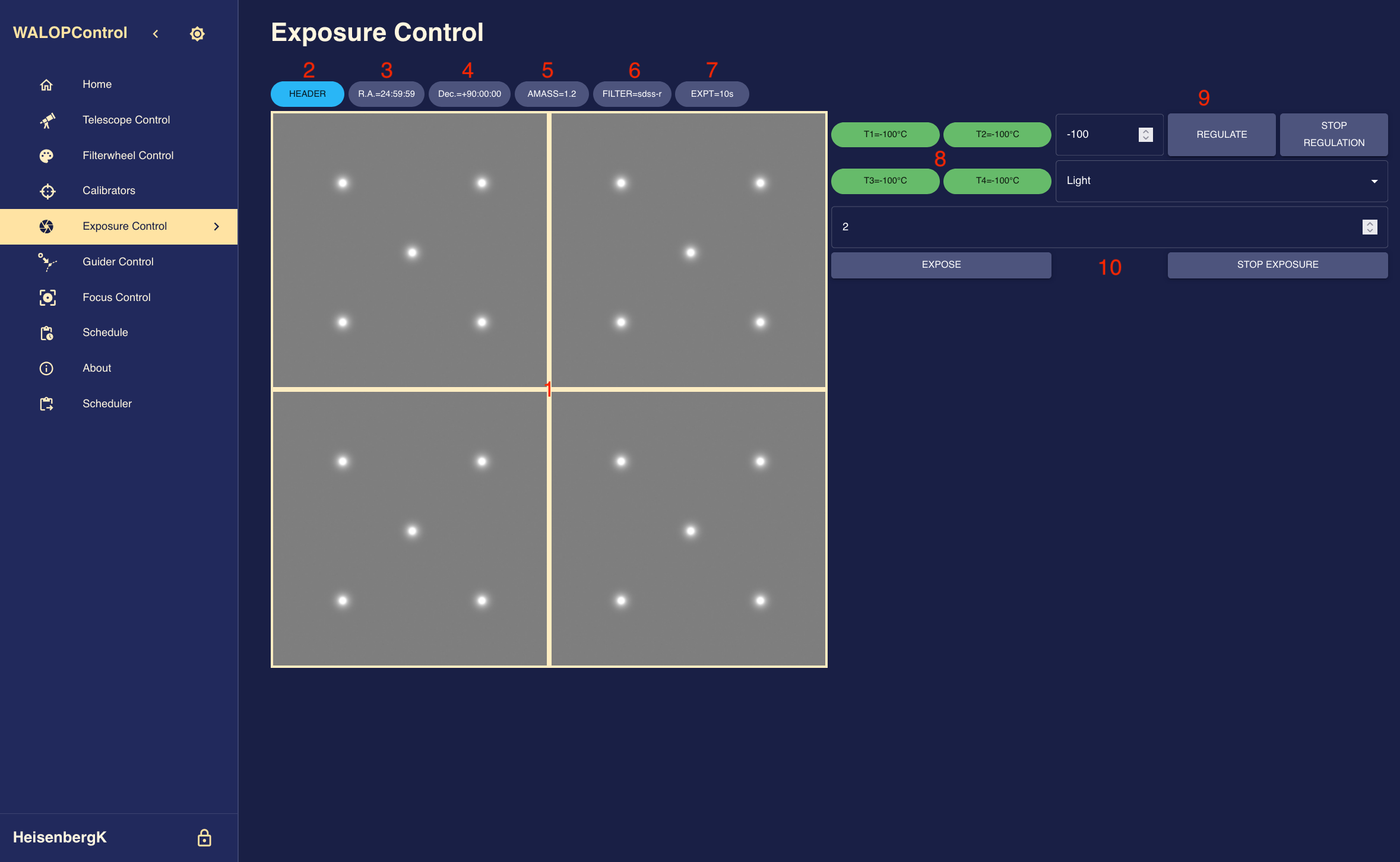}
\end{tabular}
\end{center}
\caption 
{ \label{fig:gui_camera}
The camera control page of the WALOPControl GUI. 1: 4-CCD view of the last exposure, 2: toggle a popup with the header of the displayed exposure, 3: pointing R.A. of the displayed exposure, 4: pointing Dec. of the displayed exposure, 5: pointing Airmass of the displayed exposure, 6: filter used for the exposure, 7: exposure time of the displayed exposure, 8: 4-CCD-temperatures display, 9: 4-CCD-temperature control, 10: exposure control. }
\end{figure}

\begin{figure}[!ht]
\begin{center}
\begin{tabular}{c}
\includegraphics[width=0.95\textwidth]{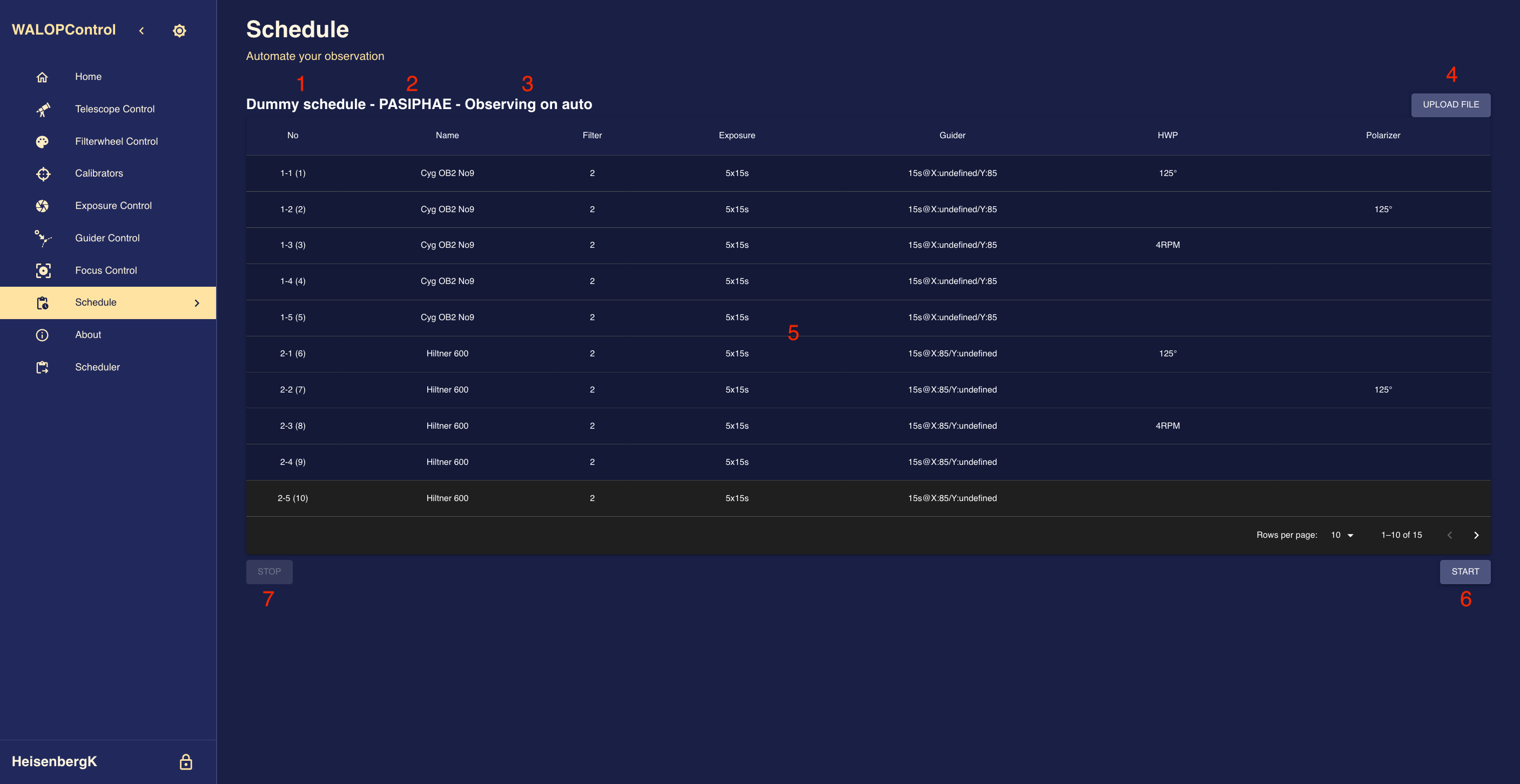}
\end{tabular}
\end{center}
\caption 
{ \label{fig:gui_schedule}
The schedule control page of the WALOPControl GUI. 1: schedule name, 2: schedule project, 3: schedule mode (auto/manual), 4: upload schedule as JSON file, 5: schedule display (blue background means target observed, light blue background means target is being observed, dark background means future target), 6: start automatic observations (if auto schedule), 7: stop automatic observations. }
\end{figure}

\begin{figure}[!ht]
\begin{center}
\begin{tabular}{c}
\includegraphics[width=0.95\textwidth]{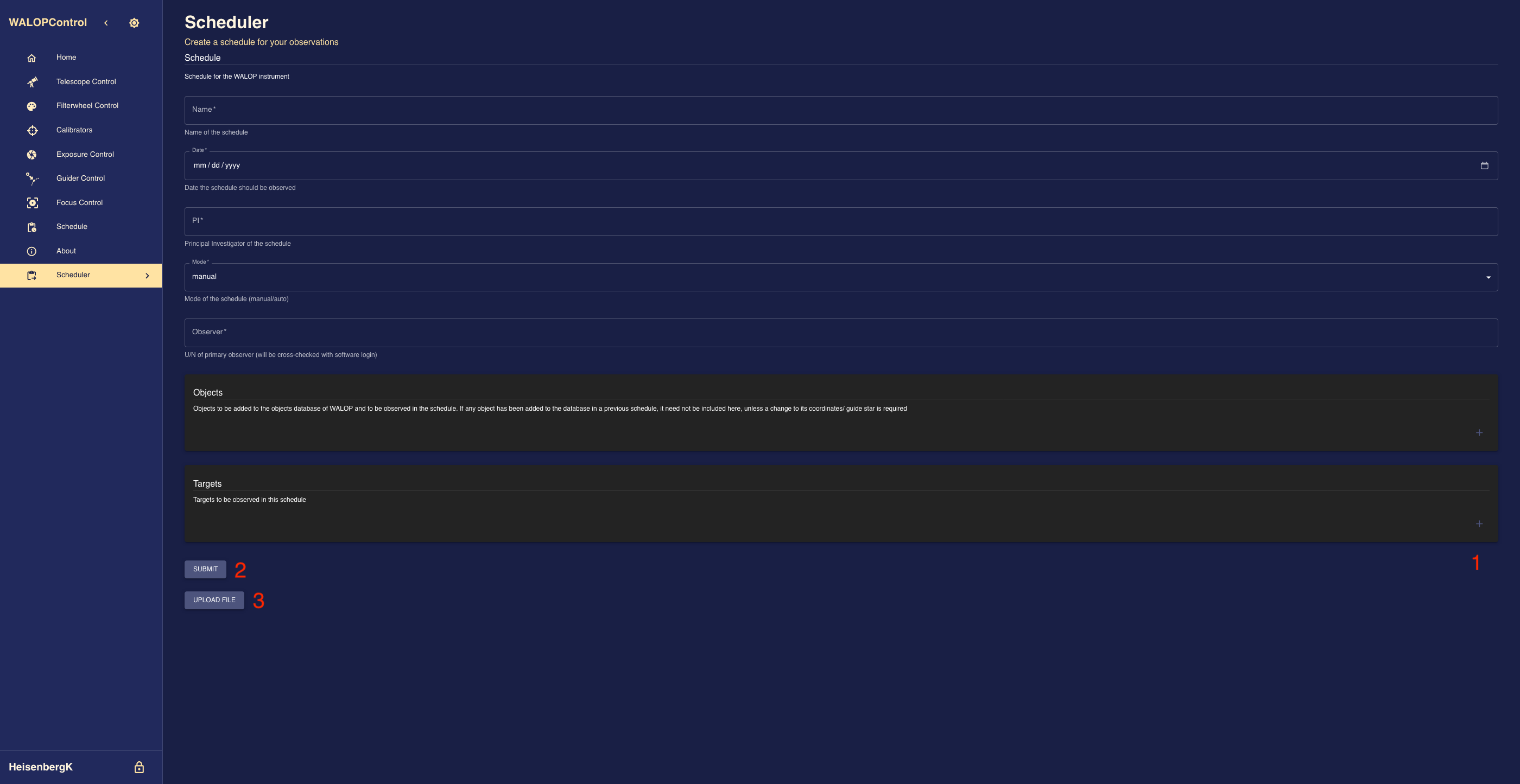}
\end{tabular}
\end{center}
\caption 
{ \label{fig:gui_scheduler}
The edit schedule page of the WALOPControl GUI. 1: the schedule edit form, 2: download schedule as JSON, 3: upload a schedule JSON to edit in the form. }
\end{figure}

Figures \ref{fig:gui_fw}, \ref{fig:gui_cal}, \ref{fig:gui_focus}, \ref{fig:gui_telescope}, \ref{fig:gui_camera}, \ref{fig:gui_schedule}, and \ref{fig:gui_scheduler} are screenshots of the WALOPControl GUI, demonstrating the control of the WALOP instruments.

\section{Database Schematic}\label{sect:ap_db}
\begin{figure}[!ht]
\begin{center}
\begin{tabular}{c}
\includegraphics[width=0.95\textwidth]{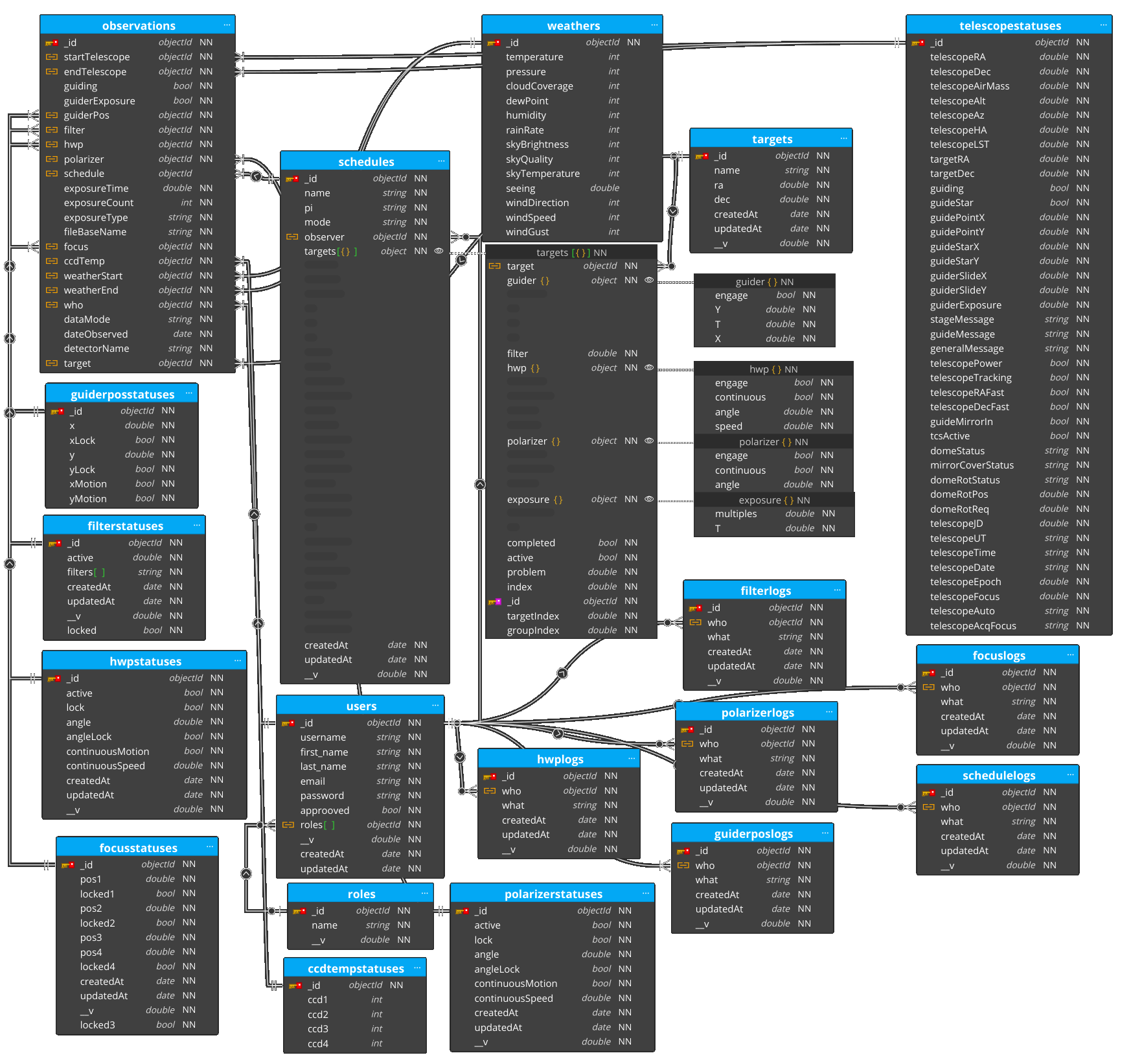}
\end{tabular}
\end{center}
\caption 
{ \label{fig:db}
The WALOPControl database.}
\end{figure}

Figure \ref{fig:db} is the schematic of the database of WALOPControl.

\section{Database List}\label{sect:ap_db_list}

WALOPControl's database has the following collections (every collection has an \_id and a \_v automatically generated attribute, containing each document's unique ID and its schema version respectively; most collections have a createdAt and updatedAt attribute containing the creation and update time of each document in them respectively):
\begin{itemize}
    \item\textbf{roles:} roles that can be assumed by WALOPControl users, has the following attributes:
    \begin{itemize}
        \item\textit{name}: the role's name (e.g. ``user", ``admin")
    \end{itemize}
    \item\textbf{users:} users of WALOPControl, has the following attributes:
    \begin{itemize}
        \item\textit{username}: the user's username
        \item\textit{first\_name}: the user's given name
        \item\textit{last\_name}: the user's family name
        \item\textit{email}: the user's email address
        \item\textit{password}: the user's password, SHA-256 hashed
        \item\textit{approoved}: whether the user is approved by an administrator
        \item\textit{roles}: the user's roles, as a list of references to the roles collection
    \end{itemize}
    \item\textbf{targets:} schedules of targets observed in sequence by WALOPControl, has the following attributes:
    \begin{itemize}
        \item\textit{name}: the targets's name
        \item\textit{pi}: the target's Right Ascension
        \item\textit{dec}: the target's Declination
    \end{itemize}
    \item\textbf{logs:} many different collections containing instrument logs, referring to the telescope, HWP, polarizer, filter-wheel, focus, scheduler, guider position, and focusers. They all have the following attributes:
    \begin{itemize}
        \item\textit{who}: reference to the user's document
        \item\textit{what}: string description of the request made by the user
    \end{itemize}
    \item\textbf{weathers:} weather info as reported by the weather station, has the following attributes:
    \begin{itemize}
        \item\textit{temperature}: air temperature at the observatory
        \item\textit{pressure}: air pressure at the observatory
        \item\textit{cloudCoverage}: cloud coverage at the observatory
        \item\textit{dewPoint}: dew point at the observatory
        \item\textit{humidity}: relative humidity at the observatory
        \item\textit{rainRate}: momentary rainfall rate at the observatory
        \item\textit{skyBrightness}: sky brightness at the observatory
        \item\textit{skyQuality}: sky quality index at the observatory
        \item\textit{skyTemperature}: sky temperature at the observatory
        \item\textit{seeing}: seeing at the observatory
        \item\textit{windDireciton}: wind direction at the observatory
        \item\textit{windSpeed}: wind speed at the observatory
        \item\textit{windGust}:  gust speed at the observatory
    \end{itemize}
    \item\textbf{filterstatuses:} status of the filter-wheel, has the following attributes:
    \begin{itemize}
        \item\textit{active}: active WALOP filter
        \item\textit{filters}: list of installed filters
        \item\textit{locked}: whether the filter is locked in position (filter-wheel not moving)
    \end{itemize}
    \item\textbf{hwpstatuses:} status of the calibration HWP, has the following attributes:
    \begin{itemize}
        \item\textit{active}: whether the HWP is in the beam
        \item\textit{lock}: whether the HWP is locked in position (not moving in/out of the beam)
        \item\textit{angle}: HWP angle (undefined if in constant rotation)
        \item\textit{angleLock}: whether the HWP angle is locked (not rotating)
        \item\textit{continuousMotion}: whether the HWP is in continuous rotation
        \item\textit{continuousSpeed}: speed of continuous rotation of the HWP
    \end{itemize}
    \item\textbf{polarizerstatuses:} status of the calibration polarizer, has the following attributes:
    \begin{itemize}
        \item\textit{active}: whether the polarizer is in the beam
        \item\textit{lock}: whether the polarizer is locked in position (not moving in/out of the beam)
        \item\textit{angle}: polarizer angle (undefined if in constant rotation)
        \item\textit{angleLock}: whether the polarizer angle is locked (not rotating)
        \item\textit{continuousMotion}: whether the polarizer is in continuous rotation
        \item\textit{continuousSpeed}: speed of continuous rotation of the polarizer
    \end{itemize}
    \item\textbf{focusstatuses:} status of the focusers, has the following attributes:
    \begin{itemize}
        \item\textit{pos1}: position of the focuser of CCD-1
        \item\textit{locked1}: whether the focuser of CCD-1 is locked in place
        item\textit{pos2}: position of the focuser of CCD-2
        \item\textit{locked2}: whether the focuser of CCD-2 is locked in place
        item\textit{pos3}: position of the focuser of CCD-3
        \item\textit{locked3}: whether the focuser of CCD-3 is locked in place
        item\textit{pos4}: position of the focuser of CCD-4
        \item\textit{locked4}: whether the focuser of CCD-4 is locked in place
    \end{itemize}
    \item\textbf{guiderposstatuses:} status of the guider positioning, has the following attributes:
    \begin{itemize}
        \item\textit{x}: x positioning of the guider around the science field of view
        \item\textit{xLock}: whether the x positioning motor of guider is locked in place
        item\textit{y}: y positioning of the guider around the science field of view
        \item\textit{yLock}: whether the y positioning motor of the guider is locked in place
        item\textit{xMotion}: opposite of xLock
        \item\textit{yMotion}: opposite of yLock
    \end{itemize}
    \item\textbf{ccdtempstatuses:} status of the CCD temperatures, has the following attributes:
    \begin{itemize}
        \item\textit{ccd1}: temperature of CCD-1
        \item\textit{ccd2}: temperature of CCD-2
        \item\textit{ccd3}: temperature of CCD-3
        \item\textit{ccd4}: temperature of CCD-4
    \end{itemize}
    \item\textbf{telescopestatuses:} status reported by the TCS, including as extra the next target to slew, has the following attributes:
    \begin{itemize}
        \item\textit{telescopeRA}: the telescope pointing Right Ascension
        \item\textit{telescopeDec}: the telescope pointing Declination
        \item\textit{telescopeAirMass}: the telescope pointing $sec(\zeta)$
        \item\textit{telescopeAlt}: the telescope pointing Altitude
        \item\textit{telescopeAz}: the telescope pointing Azimuth
        \item\textit{telescopeHA}: the telescope pointing Hour Angle
        \item\textit{telescopeLST}: the telescope Local Sidereal Time
        \item\textit{targetRA}: the next target's Right Ascension
        \item\textit{targetDec}: the next target's Declination
        \item... Other attributes about the telescope's status ...
    \end{itemize}
    \item\textbf{observations:} log and status of observations, has the following attributes:
    \begin{itemize}
        \item\textit{startTelescope}: reference to the ``telescopestatuses" document at the time of exposure start
        \item\textit{endTelescope}: reference to the ``telescopestatuses" document at the time of exposure end
        \item\textit{guiding}: whether guiding was engaged for this observation
        \item\textit{guiderExposure}: exposure time for the guider if guiding
        \item\textit{guiderPos}: reference to the ``guiderposstatuses" document at the time of exposure start (not expected to change during exposure)
        \item\textit{filter}: reference to the ``filterstatuses" document at the time of exposure start (not expected to change during exposure)
        \item\textit{hwp}: reference to the ``hwpstatuses" document at the time of exposure start (not expected to change during exposure)
        \item\textit{polarizer}: reference to the ``polarizerstatuses" document at the time of exposure start (not expected to change during exposure)
        \item\textit{schedule}: reference to the ``schedules" document, if observation is part of a schedule
        \item\textit{exposureTime}: exposure time
        \item\textit{exposureCount}: number of observations in this observation's set
        \item\textit{exposureNumber}: observation's serial number in its set
        \item\textit{exposureType}: type of exposure (light, bias, flat, calibration, dark)
        \item\textit{focus}: reference to the ``focusstatuses" document at the time of exposure start (not expected to change during exposure)
        \item\textit{ccdTemp}: reference to the ``ccdtempstatuses" document at the time of exposure start (not expected to change during exposure)
        \item\textit{startWeather}: reference to the ``weathers" document at the time of exposure start
        \item\textit{endWeather}: reference to the ``weathers" document at the time of exposure end
        \item\textit{who}: reference to the ``users" document of the observer
        \item\textit{target}: reference to the ``targets" document of the observed target
    \end{itemize}
    \item\textbf{schedules:} log and status of schedules of automated observations, has the following attributes:
    \begin{itemize}
        \item\textit{name}: name of the schedule
        \item\textit{pi}: principal investigator (PI) for the observations in the schedule
        \item\textit{pi}: observation mode for the schedule (auto or manual)
        \item\textit{observer}: reference to the ``user" document of the observer
        \item\textit{targets}: the target list of the schedule, as a list of objects with the following attributes:
        \begin{itemize}
            \item \textit{target}: reference to the ``targets" document of the target
            \item \textit{guider}: object describing whether to engage the guider for this observation, the intended guider exposure time and the guider positioning
            \item \textit{filter}: the filter to be use for this observation (1-4)
            \item \textit{hwp}: object describing the calibration HWP settings to be used for this observation (in/out of the beam, angle, or continuous rotation)
            \item \textit{polarizer}: object describing the calibration polarizer settings to be used for this observation (in/out of the beam, angle, or continuous rotation)
            \item \textit{exposure}: object describing the exposure settings for this observation (multiples and exposure time)
            \item \textit{completed}: whether the target has been observed already as part of this schedule (to be skipped if the schedule is activated)
            \item \textit{active}: whether the target is being observed as part of this schedule (to be engaged again if the schedule is reactivated)
            \item \textit{problem}: problems detected during observation of this target (to be skipped if the schedule is activated/reactivated) - 0 means no problems, other integers mean specific problems
            \item \textit{index}: the order of the target in the schedule
        \end{itemize}
    \end{itemize}
\end{itemize}

\subsection*{Disclosures}
The authors declare there are no financial interests, commercial affiliations, or other potential conflicts of interest that have influenced the objectivity of this research or the writing of this paper.

\subsection* {Code, Data, and Materials Availability} 
WALOPControl is a proprietary software developed by and under the ownership of the Institute of Astrophysics, Foundation for Research and Technology Hellas. The software is not publicly available at the moment for copyright and security reasons. For more information, please contact the corresponding author.

\subsection* {Acknowledgments}
\todo{We would like to deeply thank the anonymous reviewers for their constructive comments and suggestions that significantly improved the quality and clarity of this manuscript. Their careful reading, insightful feedback, and detailed recommendations led to substantial enhancements in both the technical content and the overall presentation. The authors are grateful for the time and effort the reviewers dedicated to this work, which has contributed greatly to its final form.}

JAK acknowledges support by means of being funded by the European Union ERC-2022-STG - BOOTES - 101076343. Views and opinions expressed are however those of the author(s) only and do not necessarily reflect those of the European Union or the European Research Council Executive Agency. Neither the European Union nor the granting authority can be held responsible for them. 

The PASIPHAE program is supported by grants from the European Research Council (ERC) under grant agreements No. 771282 and No. 772253; by the National Science Foundation (NSF) award AST-2109127;  by the National Research Foundation of South Africa under the National Equipment Programme; by the Stavros Niarchos Foundation under grant PASIPHAE; and by the Infosys Foundation.

VPa acknowledges support by the Hellenic Foundation for Research and Innovation (H.F.R.I.) under the “First Call for H.F.R.I. Research Projects to support Faculty members and Researchers and the procurement of high-cost research equipment grant” (Project 1552 CIRCE).

VPa acknowledges support from the Foundation for Research and Technology - Hellas Synergy Grants Program through project MagMASim, jointly implemented by the Institute of Astrophysics and the Institute of Applied and Computational Mathematics.

KT and AP acknowledge support from the Foundation for Research and Technology - Hellas Synergy Grants Program through project POLAR, jointly implemented by the Institute of Astrophysics and the Institute of Computer Science.

TG is grateful to the Inter-University Centre for Astronomy and Astrophysics (IUCAA), Pune, India for providing the Associateship programme under which part of this work was carried out.

VPe acknowledges funding from a Marie Curie Action of the European Union (grant agreement No. 101107047).

RS acknowledges support for this work, provided by NASA through the NASA Hubble Fellowship grant HST-HF2-51566.001 awarded by the Space Telescope Science Institute, which is operated by the Association of Universities for Research in Astronomy, Inc., for NASA, under contract NAS5-26555.


\bibliography{report}   
\bibliographystyle{spiejour}   


\vspace{2ex}\noindent\textbf{John Andrew Kypriotakis} is a postdoctoral researcher at the University of Crete, Greece, Department of Physics and at the Institute for Astrophysics of the Foundation for Research and Technology Hellas, Greece. He received his B.Sc. in Physics from University of Crete, Greece in 2017 and his PhD from the same department in 2025. He is currently working on the design of the WALOP instruments for the PASIPHAE survey and he is the sole developer/sysadmin of WALOPControl. His areas of interest are Instrumentation (incl. Software), Data Analysis and Machine Learning.

\vspace{1ex}
\noindent Biographies and photographs of the other authors are not available.

\listoffigures
\listoftables

\end{spacing}
\end{document}